\documentclass{emulateapj}
\usepackage{natbib}
\usepackage{lscape}
\usepackage{amsmath}

\slugcomment{\footnotesize \today}

\newcommand{\degree}{\mbox{$^{\circ}$}}

\newcommand{\am}{\mbox{\arcmin}}
\newcommand{\as}{\mbox{\arcsec}}

\newcommand\cmv{\mbox{cm$^{-3}$}}
\newcommand\cmc{\mbox{cm$^{-2}$}}

\newcommand{\um}{$\mu$m}
\newcommand{\lsun}{\mbox{L$_\odot$}}% Lsun
\newcommand{\msun}{\mbox{M$_\odot$}}% Msun

\newcommand{\tk}{\mbox{$T_K$}}
\newcommand{\td}{\mbox{$T_D$}}
\newcommand{\n}{\mbox{$n$}}

\newcommand{\mean}[1]{\mbox{$\langle#1\rangle$}} %generic mean for defined qu.
\newcommand{\miso}{\mbox{$M_{iso}$}} % isothermal mass

\newcommand{\ammonia}{\mbox{{\rm NH}$_3$}}

\input{epsf}

\newcommand{\Snu}{\mbox{$S_{\nu}$}}

\newcommand{\microns}{\,$\mu$m}
\newcommand{\degrees}{$^{\circ}$}
\newcommand{\sqdeg}{\,deg$^2$}
\newcommand{\percent}{\,\%}

\begin{document}

%%%%%%%%%%%%%%%%%% title %%%%%%%%%%%%%%%%%%%%%%%%%%%%%%%%%%%%%%%%
\title {\bf A Mid-Infrared Census of Star Formation Activity in Bolocam Galactic Plane Survey Sources}
\author{Miranda K.~Dunham\altaffilmark{1,2,3},
Thomas P.~Robitaille\altaffilmark{4,5},
Neal J.~Evans II\altaffilmark{2},
Wayne M.~Schlingman\altaffilmark{6},
Claudia J.~Cyganowski\altaffilmark{7,8},
James Urquhart\altaffilmark{9}}
\altaffiltext{1}{Department of Astronomy, Yale University, P.O. Box 208101, New Haven, CT 06520-8101} 
\altaffiltext{2}{Department of Astronomy, The University of Texas at Austin,
       1 University Station C1400, Austin, Texas 78712--0259}
\altaffiltext{3}{Email:  miranda.dunham@yale.edu}
\altaffiltext{4}{Tinsley Visiting Scholar, Department of Astronomy, The University of Texas at Austin, 1 University Station C1400, Austin, Texas 78712--0259}
\altaffiltext{5}{Spitzer Postdoctoral Fellow, Harvard-Smithsonian Center for Astrophysics, 60 Garden Street, Cambridge, MA 02138}
\altaffiltext{6}{Steward Observatory, University of Arizona, 933 North Cherry Ave., Tucson, AZ 85721}
\altaffiltext{7}{NSF Astronomy and Astrophysics Postdoctoral Fellow, Harvard-Smithsonian Center for Astrophysics, Cambridge, MA 02138}
\altaffiltext{8}{Department of Astronomy, University of Wisconsin, Madison, WI 53706}
\altaffiltext{9}{Australia Telescope National Facility, CSIRO Astronomy and Space Science, P.O.~Box 76, Epping, NSW 1710, Australia}

%%%%%%%%%%%%%%%%%%%% Abstract %%%%%%%%%%%%%%%%%%%%%%%%%%%%%
\begin{abstract}
We present the results of a search for mid-infrared signs of
star formation activity in the 1.1 mm sources in the Bolocam Galactic
Plane Survey (BGPS).  We have correlated the BGPS catalog with available
mid-IR Galactic plane catalogs based on the 
\textit{Spitzer Space Telescope} GLIMPSE legacy survey and
the \textit{Midcourse Space Experiment} (\textit{MSX}) Galactic plane survey.
We find that 44\%\ (3,712 of 8,358) of the BGPS sources contain at 
least one mid-IR source, including 2,457 of 5,067 (49\%) 
within the area where all surveys overlap ($10\degree < \ell < 65$\degree).  
Accounting for chance alignments 
between the BGPS and mid-IR sources, we conservatively  
estimate that 20\%\ of the 
BPGS sources within the area where all surveys overlap show signs of active star formation.  
We separate the BGPS sources
into four groups based on their probability of star formation activity.  
Extended Green Objects (EGOs) and Red \textit{MSX} 
Sources (RMS) make up the highest probability group, while the lowest
probability group
is comprised of ``starless'' BGPS sources which were not matched
to any mid-IR sources.  The mean 1.1 mm flux of each group increases with increasing 
probability of active star formation.  
We also find that the ``starless'' BGPS sources are
the most compact, while the sources with the highest probability of 
star formation activity 
are on average more extended with large skirts of emission.  
A subsample of 280 BGPS sources
with known distances demonstrates that mass and
mean H$_2$ column density
also increase with probability of star formation activity.

\end{abstract}

%%%%%%%%%%%%%%%%%%% Main text %%%%%%%%%%%%%%%%%%%%%%%%%%%%

\section{Introduction}\label{intro}

The formation of massive stars remains one of the prominent puzzles
in astronomy (see review by McKee \& Ostriker 2007).  A current
goal in high-mass star formation studies is to determine an evolutionary 
sequence and lifetimes similar to the Class
system in low-mass star formation (e.g.~Shu et al.~1987).  While
the evolutionary sequence is still open to considerable debate, one
proposed sequence is summarized in recent reviews by Churchwell (2002)
and Zinnecker \& Yorke (2007).
This evolutionary sequence begins with prestellar cores or clumps: 
gravitationally bound over-densities within a molecular cloud 
which show signs of inward motion but have not yet begun to
form a protostar.  Since prestellar cores are lacking any
internal heating, their temperatures are typically 10-20 K
and their spectral energy distributions peak in the far-infrared.

Infrared dark clouds (IRDCs), seen in absorption against a diffuse
mid-infrared background, are considered the cold precursors 
to star clusters (Rathborne et al.~2006), and thousands have been cataloged 
from the \textit{MSX} and GLIMPSE surveys (Egan et al.~1998; Carey et al.~1998; 
Simon et al.~2006; Peretto \& Fuller 2009). 
While IRDCs may be the sites of the earliest stages of massive star 
formation, the clumps and cores within IRDCs can host massive young 
stellar objects (MYSOs) in various stages of evolution (e.g.~Chambers
et al.~2009; Rathborne et al.~2005, 2006, 2007, 2008, 2010). 
While a prestellar core is only seen as an IRDC if it is located
at the near distance and there is enough background IR emission 
to absorb, IRDCs are seen in emission at far-infrared,
submillimeter and millimeter wavelengths, allowing for detection and
identification of prestellar cores throughout the Galaxy.

The second stage in the proposed evolutionary sequence of Churchwell (2002)
 is hot cores:
compact (diameter $<$ 0.1 pc), dense ($n>10^7$ \cmv), and warm (T$>100$K)
molecular cloud cores with high molecular line brightness temperatures 
(Kurtz et al.~2000) which are a result of internal heating from
a massive protostar (Churchwell 2002).  These hot cores will not contain 
any radio free-free emission because the infall rate will be high enough
for the infalling material to absorb the UV flux from the central
protostar. 

Once the accretion begins to taper off, a small but detectable
HII region will form.  Thus, the third evolutionary stage is 
thought to include Hyper-Compact HII (HCHII) regions which will rapidly
expand to Ultra-Compact HII (UCHII) regions. This stage can be
characterized by the presence of radio emission (Churchwell 2002).  
The final stages are compact and classical HII regions.  The HII region has
now expanded and begun disrupting the parent molecular cloud
while revealing the embedded stellar population in the optical and
infrared.  

This evolutionary sequence based on the size of HII regions may
be too simplistic.  Recent simulations of the gravitational collapse
of a massive molecular cloud suggest that the size and morphology 
of HII regions can vary drastically during the formation of a
massive star, with the HII region expanding and contracting repeatedly
through the HCHII and HII classifications (Peters, et al.~2010a; 2010b). 
These simulations suggest that the size of an HII region will only 
depend on age late in the lifetime of the HII region once accretion has halted.

Other evolutionary sequences have been proposed based on the presence
of various types of masers, radio emission, thermal emission from
warm dust, as well as any combination or complete lack of these star 
formation tracers (e.g.~De Buizer et al.~2005; Minier et al.~2005; 
Ellingsen et al.~2007;  Longmore et al.~2007; Purcell et al.~2009; 
Breen et al.~2010).  These studies have included, at most,
on the order of a hundred sources, and while detailed studies
of small samples are crucial for understanding the results of 
large-scale surveys, Galaxy-wide samples are needed in order
to solidify an evolutionary sequence.

Recent and ongoing (sub)millimeter wavelength 
continuum surveys of the Galactic plane are
providing a novel inventory of star formation sites within the Galaxy
at unprecendented sensitivity and resolution (BGPS, Aguirre et al.~2011;
ATLASGAL, Schuller et al.~2009; JPS; Di Francesco et al.~2008).
These surveys are detecting thousands of sites of massive star formation and
providing a statistically significant sample with which to begin
studying star formation on a Galactic scale.  However, these surveys
alone cannot provide direct evidence of a source's evolutionary state and 
whether it is actively forming stars.

Fortunately, a wealth of other multiwavelength surveys that can
provide evidence of a source's evolutionary state are already 
available or currently being observed.
For example, the \textit{Spitzer Space Telescope} 
GLIMPSE survey (Benjamin et al.~2003; Churchwell et al.~2009)
at 3.6 to 8.0 \um\ can provide information regarding the population of 
embedded young stellar objects (YSOs) found throughout the Galactic plane via
mid-infrared colors (e.g. Robitaille et al.~2008)
and the presence of outflows from massive young stellar objects traced
by extended 4.5 \um\ emission (e.g. Cyganowski et al.~2008).  
Similarly, the Red \textit{MSX} Source survey (RMS; Hoare et al.~2004) has
identified a reliable sample of massive YSOs by following up color-selected
\textit{MSX} YSO candidates.
The \textit{Spitzer} MIPS Galactic Plane Survey
(MIPSGAL; Carey et al.~2009) has mapped the emission from warm dust in the
Galactic plane at 24 and 70 \um\ and can also identify embedded protostars.
The on-going Herschel Infrared Galactic Plane Survey (Hi-Gal; Molinari et al.~2010)
will map the same region as the \textit{Spitzer} surveys at 60-600 \um\ and will
identify YSOs at wavelengths spanning the peaks of their spectral energy distributions.
Together with the shorter wavelengths from \textit{Spitzer}, Hi-Gal will provide
accurate protostellar luminosities (e.g.~Elia et al.~2010).
At longer wavelengths, the Co-Ordinated Radio `N' Infrared Survey for High-mass
Star Formation (CORNISH; Purcell et al.~2008) will provide a complete catalog
of UCHII regions.
To gain the most insight from the copious Galactic plane surveys, we must
begin correlating their data sets.  

In this paper, we present the results 
of a comparison of the Bolocam Galactic Plane Survey (BGPS) 
and available, Galaxy-wide mid-IR based YSO catalogs with the goal
of characterizing the star formation activity within the BGPS sources.
This work represents the first step in determining the evolutionary 
state of the BGPS sources.  Further comparison to other Galactic plane 
surveys, such as MIPSGAL and CORNISH, is required and will be the 
subject of future papers. 
Section \ref{datasets} presents the details of the various data sets
included.  The basic results are presented in Section \ref{results}, 
including the cross-matching method, a discussion of the prevalence of
chance alignments, and the basic cross-matching statistics.  Section
\ref{discussion} discusses trends in various millimeter properties
as a function of star formation activity, and Section \ref{summary} 
provides a summary of the work.

\section{Data Sets}\label{datasets}

\subsection{The Bolocam 1.1 mm Galactic Plane Survey}\label{BGPS}
The Bolocam Galactic Plane Survey\footnote{See http://milkyway.colorado.edu/bgps/.} 
(BGPS; Aguirre et al.~2011) has surveyed 170 square degrees
of the Northern Galactic Plane in 1.1 mm continuum emission using 
the facility long-wavelength camera, Bolocam 
(Glenn et al.~2003; Haig et al.~2004), at the Caltech 
Submillimeter Observatory (CSO) on Mauna Kea.  The survey consists of two
distinct portions:  a blind survey of the inner Galaxy spanning 
$-10.5\degree < \ell < 70.0\degree$ where $|b|<0.5\degree$ everywhere
except for 1.0\degree\ sections in $\ell$\ centered at 
$\ell=3\degree, 15\degree, 30\degree, \mbox{and } 31\degree$ where $|b|<1.5\degree$, and a targeted 
outer Galaxy
portion including Cygnus-X ($70\degree < \ell < 90.5\degree$, $|b|<1.5\degree$), 
the Perseus Arm ($\ell \sim$ 111\degree), IC1396 ($\ell\sim$99\degree, 
$b$$\sim$3.5\degree), the W3/4/5 region ($\ell\sim135\degree$,$b\sim0.5\degree$),
and the Gemini OB1 Molecular Cloud ($\ell\sim190\degree$,$b\sim0.5\degree$).

Bolocam is a hexagonal array of 144 bolometers with a filter centered at 
268 GHz with an effective band center of 271.1 GHz (Aguirre et al.~2011)
and a bandwidth of 45 GHz.  The filter excludes the CO(2-1) 
emission line.  Each individual bolometer has a Gaussian beam with a 
FWHM of 31\as, and the instrument has an instantaneous field of view
of 7.5\am\ due to the separation of the bolometers in the focal plane 
of the instrument.  For more information regarding Bolocam, see
Glenn et al. (2003), and Haig et al. (2004).

A reduction pipeline was created specifically for the BGPS survey, and
is described in Aguirre et al. (2011).  In order to retain the extended
structure in the final maps, an iterative sky cleaning method was employed.
At each iteration a model of the sky emission is updated and subtracted from 
the overall data timestream to reveal a time stream containing only
the astrophysical signal.  The astrophysical signal is then mapped
with 7.2\as$\times$7.2\as\ pixels and  
the effective beam size for the BGPS is 33\as.  

Observations were obtained with scans along both $l$ and $b$ using the
raster scan mode and a scan speed of 120\as s$^{-1}$.  Sensitivity to extended
structure up to 5.9\am\ angular size was retained by not 
utilizing chopping.  The full flux density of objects larger than
5.9\am\ is not recovered, but long filaments with at least one 
dimension smaller than 5.9\am\ are still detected (Aguirre et al.
2011).  The flux density was calibrated based on observations
of planets.  Aguirre et al.~(2011) compared BGPS with 1.2 mm data
acquired with MAMBO on the IRAM 30 m (Rathborne et al.~2006; 
Motte et al.~2003, 2007) and
SIMBA on the SEST 15 m (Matthews et al.~2009) and found that the 
v1.0 BGPS data were systematically lower and require
a mean scaling factor of 1.5$\pm$0.15 to 
align the BGPS fluxes with the previous MAMBO and SIMBA data sets.  
Here we present
the v1.0 data with the scaling factor of 1.5$\pm$0.15 applied.

Millimeter continuum sources were extracted from the iterativly mapped
images using the BGPS source extraction software, Bolocat (Rosolowsky
et al.~2010).  Bolocat employs a seeded watershed method (Soille 1999)
which first identifies all pixels above a signal-to-noise ratio of 2.0,
identifies contiguous groups of pixels above this signal-to-noise ratio,
culls any groups which contain fewer than a beam's worth of pixels (22 7.2\as\ pixels),
and then expands each remaining high signal-to-noise region through a 
nearest-neighbor algorithm to include all adjacent pixels above a
signal-to-noise ratio of 1.0.  Extending the region to include pixels down to
1 $\sigma$ is based on the reasoning that marginal emission next to regions
of significant emission is likely real.  The software then breaks each region up
into smaller substructures based on the contrast between local maxima.  
Through the nearest neighbor algorithm each source is assigned a group 
of contiguous pixels and a source number.  Bolocat then produces a 
label map, in which every pixel assigned to a source is 
assigned the source number.  Source properties are then computed
via emission-weighted moments.  All flux dependent properties presented
in this work are based on the integrated flux density which is simply
the sum of all pixels assigned to a BGPS source.  This flux density
avoids contamination from nearby sources that can occur with aperture
photometry.  For more details regarding source 
extraction and parameter estimation, see Rosolowsky et al.~(2010).  

In the generation of the Bolocam catalog (Rosolowsky et al., 2010) 
we evaluated both the completeness limit (false negatives as a function of 
signal-to-noise ratio) and the spurious detection incidence (false positives 
as a function of catalog parameters).  The completeness study was presented 
as part of the original catalog manuscript.  The catalog parameters were 
specifically chosen to minimize false detections as well as false negatives.  
To evaluate the false-positive incidence, we inserted an array of 2600 false 
sources into the data time stream for a blank field of observation.  The time 
stream was for the l=111 field which had been cleaned of signal as a byproduct 
of the iterative mapping procedure (Aguirre et al. 2011).  The injected sources 
had a range of brightnesses and the recovery fraction as a function of 
brightness established the completeness limit.   The algorithm recovered 0 
false detections -- all sources in the final catalog were the injected sources.  
While this may be surprising, the catalog's relatively high effective 
completeness limit of $5\sigma$ is a byproduct of eliminating false detections.  
Taking this as $0\pm 1$ over the 4 deg$^2$ area of the field leads to an upper 
limit of $\sim 40$ false detections in the survey, or 0.5\%.  We note that the 
injected sources were point sources and extended source recovery tests found 
that the BGPS catalog software often broke up extended emission into multiple 
objects, but it did not lead to the false detection of sources not associated 
with emission.  Based on this test, we believe the BGPS catalog 
contains only a very small number of false sources.

The BGPS catalog is $>$99\%\ complete at the 5$\sigma$\ level.  While
$\sigma$\ varies across the survey due to the varying number of observations
and observing conditions, the survey as a whole is $>$98\%\ complete
at the 0.4 Jy level (see Figure 9 in Rosolowsky et al.~2010).  
Assuming a dust temperature of 20 K (Dunham et al.~2010), this 
completeness level translates into a minimum mass of
5 \msun, 21 \msun, 130 \msun, 524 \msun, and 750 \msun\ at 1 kpc, 2 kpc, 
5 kpc, 10 kpc, and 12 kpc, respectively.  
However, since the BGPS filters out diffuse
emission on scales larger than 5.9\am\ (Aguirre et al.~2011), the types
of sources the BGPS is sensitive to depends on distance (see Section 7.2 of 
Dunham et al.~2010 and Section 6 of Rosolowsky et al.~2010).
At the nearest distances (up to approximately 2 kpc), the 
BGPS will be sensitive to cores 
(mass M$=0.5-5$ \msun, volume-averaged density n$=10^4-10^5$ \cmv, and physical radius R$=0.01-0.1$ pc; Bergin \& Tafalla 2007),
while at intermediate distances the BGPS will be sensitive to clumps
(M$=50-500$ \msun, n$=10^3-10^4$ \cmv, R$=0.15-1.5$ pc; Bergin \& Tafalla 2007),
and at the furthest distances BGPS will detect 
entire molecular clouds (M$=10^3-10^4$ \msun, n$=50-500$ \cmv, R$=1-7.5$ pc; 
Bergin \& Tafalla 2007).

The BGPS images and catalog have been released to the public and are
hosted by the Infrared Processing and Analysis Center via the
NASA/IPAC Infrared Science 
Archive\footnote{See http://irsa.ipac.caltech.edu/data/BOLOCAM\_GPS/}
(IPAC).

\subsection{GLIMPSE Red Source Catalog}\label{redsources}

In addition, we made use of the GLIMPSE survey, a \textit{Spitzer}/IRAC Legacy survey of the Galactic mid-plane (Benjamin et al.~2003; Churchwell et al.~2009) at 3.6, 4.5, 5.8, and 8.0\microns. The GLIMPSE I survey covers $10$\degrees\ $\le|\ell|\le65$\degrees\ and $|b|\le1$\degrees\, and the GLIMPSE~II survey fills in the region for $|\ell|<10$\degrees\, with $|b|\le1$\degrees\ for $|\ell|>5$\degrees, $|b|\le1.5$\degrees\ for 2\degrees$<|\ell|\le5$\degrees, and $|b|\le2$\degrees\ for $|\ell|\le2$\degrees. The total area of the GLIMPSE I and II surveys is 274\sqdeg, and the overlap region with the BGPS survey is $-10.5$\degrees$<\ell<65$\degrees, $|b|<0.5$\degrees\ and $|b| < 1.0\degree\ \mbox{at}\ \ell=$3\degrees, 15\degrees, 30\degrees, and 31\degrees. The data products include both highly reliable Catalogs, and less reliable but more complete Archives.

Robitaille et al.~(2008; hereafter R08) compiled a highly reliable census of 18,949 GLIMPSE point sources with excess emission at mid-infrared wavelengths, of which 11,649 were classified as candidate young stellar objects (YSOs), and 7,300 as candidate asymptotic giant branch (AGB) stars. The AGB classification was further separated into two categories consisting of ``standard'' AGB stars and ``extreme'' AGB stars.  The ``extreme'' AGB stars have high mass-loss rates and therefore significant circumstellar dust.  In this paper, we refer to the ``standard'' and ``extreme'' AGB stars as sAGB and xAGB stars, respectively, and both types together as simply AGB stars.  The separation of the sources into YSOs and AGB stars was based on IRAC and MIPS color and magnitude selection criteria, and was only approximate. In reality, $50-70$\percent\ of the sources are likely to be YSOs, and $30-50$\percent\ are likely to be AGB stars.

The R08 census was designed to maximize reliability. As the sensitivity to point sources decreases in areas of bright diffuse emission -- which are common at 8\microns\ -- the census used stringent magnitude selection criteria so as not to be affected by this. As a result, the R08 census may be missing fainter YSOs that are still detected by GLIMPSE. Therefore, for the current work, we compile a list of red GLIMPSE sources that were missed in the R08 census, with the caveat that this complementary list is therefore less reliable, and is background rather than flux limited, especially at 8\microns.

We constructed this complementary sample as follows. We first increased the reliability of the GLIMPSE catalog fluxes in a similar way to R08. First, each position in the GLIMPSE survey was observed at least twice, and we therefore removed fluxes from the Catalog that depend only on a single detection. We required the fractional rms of the individual detections and the fractional error in the combined flux to be less than 15\percent, otherwise the flux was removed. We then rejected sources that were not detected at least at 4.5 and 8.0\microns, and finally we selected the red sources by requiring $[4.5]-[8.0]>0.75$, a less stringent color selection than that used for R08 ($[4.5]-[8.0]>1.0$). This results in 64,859 sources being selected, excluding the R08 sources, which we refer to as \textit{additional GLIMPSE sources}.

Robitaille et al.~(2008) estimated that the red source catalog is 65.7\%\ complete.  
Additionally, they explored the range of luminosities
of YSOs detectable as a function of distance.  The sensitivity of the GLIMPSE survey results
in a lower flux limit, and Robitaille et al.~imposed an upper flux limit in order to exclude
point sources with unreliable fluxes due to saturation.
They used the models of Robitaille et al.~(2006) to calculate average 
$[4.5]$ and $[8.0]$ magnitudes for Stage I and II YSOs as a function of luminosity,
and calculated the range of distances at which a YSO would be included in their Red
Source Catalog.  A 1 \lsun\ YSO within 1 kpc
will be included in the red source catalog, while a 10 \lsun\ YSO will be included if it 
is between 0.2 and 2 kpc.  A $10^3$ \lsun\ YSO between 2 and 10 kpc will be included in the
R08 catalog, and a $10^4$ \lsun\ YSO between 4 and 11 kpc will also be included. 
The most luminous YSOs will only be included at the farthest distances due to the upper flux
limit imposed to avoid saturation, while the lowest luminosity sources will only be found at the 
nearest distances due to the sensitivity limits of the GLIMPSE survey
(see Figure 19 and Section 4.5 of Robitaille et al.~2008).

\subsection{Extended Green Objects Catalog}\label{egos}

The GLIMPSE Catalogs and Archives are point source catalogs, and therefore exclude any extended YSOs. In particular, massive embedded YSOs driving outflows can in some cases be seen as extended sources at 4.5\microns, as this band contains H$_2$  emission lines as well as CO band heads (Reach et al.~2006, Smith et al.~2006). These lines can be excited by shocks, which are expected to be present where outflows interact with the interstellar medium. Cyganowski et al.~(2008) compiled a list of 304 such objects, dubbed Extended Green Objects\footnote{The `green' adjective comes from the fact that these objects appear as green in IRAC three color images, since these often use 4.5\microns\ for the green channel} (EGOs), from the GLIMPSE I survey.  The EGOs are likely tracing outflows from \textit{massive} YSOs since they show a strong correlation with IRDCs and Class II methanol masers (Cyganowski et al.~2009), both of which are associated with early stages of massive star formation (Cyganowski et al.~2008).  The overlap area with the BGPS survey is $10$\degrees$<\ell<65$\degrees, $|b|<0.5$\degrees and $|b|<1.0$\degrees\ at $\ell =$15\degrees, 30\degrees, and 31\degrees.

The EGO catalog is certainly not a complete census of YSOs with shocked outflows. 
The requirement of extended 4.5 \um\ emission likely resulted in the preferential inclusion
of YSOs at relatively small distances, as the angular extent decreases with distance. 
Also, as noted in Cyganowski et al.~(2008), EGOs near bright sources or in areas with 
significant PAH emission and diffuse green emission extended on scales less than 10\as\
are likely absent from the catalog.  

Thus, the EGO catalog may include more massive, 
higher luminosity, nearby sources which were excluded from the R08 catalog because of
the saturation limits.

\subsection{Red \textit{MSX} Source Catalog}\label{rms}

The Red \textit{MSX} Source (RMS) 
survey\footnote{See http://www.ast.leeds.ac.uk/RMS} 
(Hoare et al.~2004; Urquhart et al.~2008b) was designed to 
follow-up and confirm massive young stellar objects (MYSO) candidates selected from the 
Midcourse Space Experiment (\textit{MSX}) Galactic plane survey 
(Price et al.~2001). The Galactic plane survey included all longitudes 
from 0 to 360\degrees, and latitudes within $|b|<5$\degrees\ and 
consisted of imaging at 8.28, 12.13, 14.65, and 21.3\microns\ with 
a resolution of 18\arcsec\ at all wavelengths (Price et al.~2001). 
A total of 1,992 MYSO candidates were identified using 
color selection criteria.  Low- and intermediate-mass YSOs were excluded from
the catalog.  All MYSO candidates were subsequently 
followed-up to confirm their nature (Mottram et al.~2007; Urquhart et 
al.~2007a, 2007b, 2008a, 2009a, 2009b) and were
classified into the following categories:  ``Evolved Star'', 
``PN'', ``OH/IR Star'', 
``Young/Old? Star'', `` HII Regions'', ``HII/YSO'', and ``YSO''.
In this work we include the ``PN'', ``OH/IR Star'', and ``Evolved Star''
classifications in a general ``Evolved Star'' category.
The ``YSO/HII'' classification is assigned when both an HII Region and YSO 
are present or if a definitive classification is not possible, while the
``Young/old'' classification is assigned to sources which look like
evolved stars but have strong CO emission along their line of sight
resulting in an uncertain classification.
The BGPS overlaps the RMS survey where $\ell \ge 10\degree$.

%Figure 1
\begin{figure*}[t]
\begin{center}
\epsscale{1.0}
\plotone{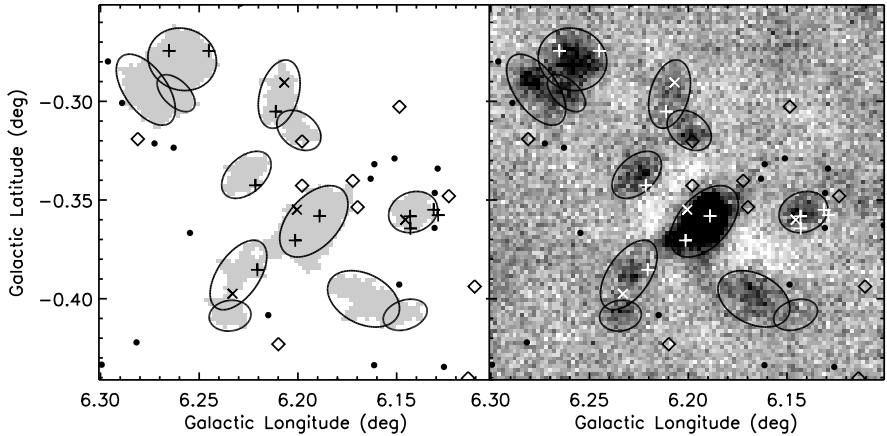} 
\figcaption{\label{labelmap} 
 Left:  The gray scale image shows the label map from a small portion of the BGPS.  Right:  The corresponding BGPS image shown in inverted grayscale ranging from 0.2 to -0.1 Jy/beam.  In both images, the black ellipses mark the positions and sizes of the extracted BGPS sources as determined from emission weighted moments.  Crosses and diamonds represent GLIMPSE R08 red sources which are and are not associated with a BGPS source, respectively. Plus signs and circles mark additional GLIMPSE sources which were matched and not matched with a BGPS source, respectively.  Note the BGPS source near $(\ell,b)\sim(6.23$\degrees$,-0.39$\degrees$)$.  The ellipse encloses many pixels which do not include millimeter emission, thus demonstrating the benefit of using the label map for matching rather than the mean source properties which are shown here as the black ellipses.  
}
\end{center}
\end{figure*}

Unlike the R08 sample, the RMS selection criteria did not include an upper-limit
on fluxes, and as such the highest luminosity sources at all distances
will be included in the catalog.  YSOs with L$\sim10^3$ \lsun\ will be included
to 3 kpc, while $10^4$ \lsun\ YSOs will be included to a distance of 
10 kpc, and $10^5$ \lsun\ YSOs will be included across the entire Galaxy
(see Figure 2 of Urquhart et al.~2008b).  The luminosity range of the RMS
survey overlaps with the range in the R08 red source catalog, and fills
in the region of luminosity space excluded from the R08 catalog because
of the saturation limitations.

\section{Results}\label{results}

\subsection{Cross-Matching Method}\label{matchingmethod}
The data sets were cross-matched based on spatial coincidence.
All infrared sources located within a BGPS source as defined
by the label maps (see \S\ref{BGPS}) are considered to be 
associated with the BGPS source.
The label maps produced as a part of the BGPS source extraction 
(Rosolowsky et al. 2010) represent the full extent of each 1.1 mm
continuum emission source down to the 1$\sigma$ level of the original
image.  We require that the catalog position of each mid-IR source fall
within a pixel assigned to a BGPS source.  The EGO sources
subtend more than a single BGPS pixel; however we find that
ignoring the extended emission and simply using the catalog position
corresponding to the peak of the 4.5 \um\ emission does not
adversely affect the matching statistics for the EGO catalog.
Figure \ref{labelmap} shows the label map and corresponding image
centered near $l\sim6$\degree\ with the positions of the matched 
(crosses, plus signs) and unmatched (diamonds, circles)
R08 and additional GLIMPSE sources, respectively.  The catalog parameters 
for each BGPS source (major
and minor axes and position angle) are displayed
as the black ellipses.  The BGPS sources clearly do not fill the
black ellipses, highlighting the benefit of using the label maps rather 
than the BGPS mean source properties when cross-matching catalogs.

\subsection{Chance Alignments}\label{chancealignments}
In matching various catalogs of mid-IR sources to the BGPS sources, we need to take into account the possibility of chance alignment of mid-IR sources with BGPS clumps. Since we do not have distance estimates to the mid-IR sources or most BGPS sources, we cannot determine which mid-IR sources that lie along the same line of sight as a BGPS source are in fact chance alignments. Instead, we can estimate the fraction of chance alignments statistically.

In order to do this, we first consider
the fraction of the BGPS area that contains mm emission and 
assume that the mid-IR sources are evenly 
distributed across the survey area.  In this case, the fraction 
of area covered with mm emission corresponds to the fraction of
mid-IR sources that are chance alignments with the mm continuum
emission.  We calculate this fraction for the region that is common 
to all surveys:  $10\degree < \ell < 65\degree$, and $|b| \leq 0.5\degree$ 
with extensions to $|b| \leq 1.0\degree$ as described in \S\ref{BGPS}.
We find that 2.1\%\ of the BGPS area contains source emission,
and therefore 2.1\%\ of the mid-IR sources will overlap mm continuum 
emission by chance.

%Figure 2
\begin{figure*}
\epsscale{0.9}
\plotone{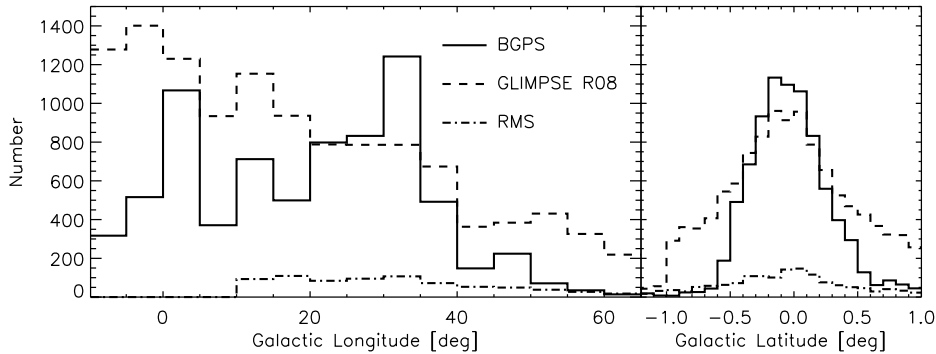} 
\figcaption{\label{histlb} 
 The distribution in Galactic longitude (left) and Galactic latitude (right)
of the BGPS sources (solid line), GLIMPSE Red Sources with $\ell \ge -10$\degree\ (dashed), and all RMS sources (dash-dot).  The EGO catalog is not plotted due to the small number of sources (84).  The sources are clearly concentrated toward the Galactic mid-plane ($b\sim 0$\degree), and also show structure in Galactic longitude.
}
\end{figure*}

\begin{figure*}
\epsscale{0.9}
\plotone{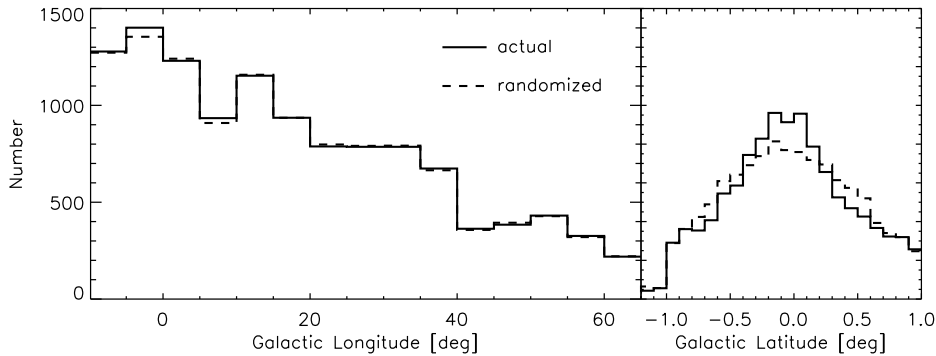}
\figcaption{\label{histlb_glimr08ran}
Distributions of Galactic longitude (left) and latitude (right) of GLIMPSE red sources with $\ell \ge -10$\degree\ for the original (solid lines) and randomized (dashed lines) catalogs.  Randomizing by 1\degree\ has very little effect on the distribution of $\ell$, but does slightly wash out the peak in the distribution of $b$.  Overall, the trend of more sources located near the Galactic mid-plane ($b\sim0$\degree) is maintained.
}
\end{figure*}

In reality, the distribution of the mid-IR sources is not even across the survey areas, but is concentrated towards the Galactic mid-plane, and also decreases with increasing longitude (Figure \ref{histlb}).  Furthermore, even a population of mid-IR sources unrelated to star formation could display a similar non-uniform spatial distribution.  To obtain a more realistic estimate of chance alignments, we attempt to preserve the large-scale distribution of the mid-IR sources by duplicating the catalogs and randomizing each source's coordinates within a circle of radius 1\degree\ (in cases where mid-IR sources fell out of their original survey area, their positions were re-randomized).  In this way, any population unrelated to star formation would preserve its large-scale distribution and would not show a difference in the number of chance alignments, while any star forming population that was originally preferentially aligned with BGPS emission on arcsecond to arcminute scales would lose such a correlation. 
Figure \ref{histlb_glimr08ran} shows the distributions of Galactic longitude (left) and latitude (right)
for the GLIMPSE red sources with $\ell \ge -10$\degree\ for the original (solid lines) and 
randomized (dashed lines) catalogs.  Overall, the trends in the distributions of the original
catalog are preserved in the randomized catalog although the peak in sources toward $b\sim0$\degree\
is slightly decreased.  To investigate the effect of the randomization radius used, we repeated the experiment with a radius of 0.1\degree\ and found that the fraction of chance alignments increases slightly.  All GLIMPSE R08 source types (xAGB, sAGB, and YSO) taken together have
8.1\%\ chance alignments with a 1\degree\ randomization.  If we use only a 0.1\degree\ randomization,
we expect 10.7\%\ chance alignments.  The choice of radius for the randomization does not
severely alter the chance alignment results.

%Figure 3

Multiple randomized catalogs were created for each mid-IR data set
in the interest of obtaining an error bar on the number of
chance alignments. For the GLIMPSE catalogs, we randomized the catalogs 10 times, while for the EGO and RMS catalogs we randomized the catalogs 100 times to avoid issues due to small-number statistics.  Column 5 of Table 1 presents the
percent of matched mid-IR source types that are expected to be chance 
alignments, and column 9 shows the percentage of BGPS sources expected to be
matched due to chance alignments with each mid-IR source type.   Overall, around 7\% of mid-IR sources would be expected to match the BGPS emission by chance alignments.

\subsection{Matching Statistics of the mid-IR sources}\label{matchstatssection}

Columns 2-5 of Table \ref{matchstats} present a summary of the matching
statistics for each mid-IR catalog, including both the number and fraction of mid-IR sources that are matched to BGPS emission. 
The EGOs are most often matched to BGPS emission, with 90\% of EGO sources matched (more than ten times that expected from random chance alignment). In comparison, the RMS sources have a lower overall matched fraction (68\%). However, if we consider the different types of RMS sources separately, it becomes apparent that the types of RMS sources associated
with star formation (e.g. HII Regions, HII/YSO, and YSO) have a much higher fraction of matches with BGPS sources. Table \ref{matchstats} lists the matching statistics
for the RMS sources by category.  
The Young/old star, HII Region, HII/YSO, and YSO categories have matched fractions approximately $8-10$ 
times the fraction expected by chance alignments (80-90\%).  
In contrast, the fraction of RMS evolved stars matched with BGPS emission is perfectly 
consistent with that expected from chance alignments.  Of the 133 RMS sources classified as AGB stars within the
``Evolved stars'' category, only 12 were matched with BGPS sources.

The GLIMPSE R08 YSOs tend to be less often associated with mm emission (33\% - or 4 times that expected from chance alignments).  This low fraction 
could be due to several reasons.  
First, Stage II
YSOs, which are present in the GLIMPSE R08 catalog, no longer have an envelope and are unlikely to possess the
density and angular size required for a detection in the BGPS
(Dunham et al.~2010, \S7.2). Secondly, the separation of YSOs
and AGB stars in the GLIMPSE Red Source catalog was only
approximate, and refining the magnitude and color
selection criteria could result in a higher fraction of 
matched YSOs. 
The fraction of AGB stars from the GLIMPSE R08 catalog matched to mm emission is 
slightly above the level expected for chance alignments. This could be due to the 
approximate YSO/AGB separation which results in contamination from genuine YSOs 
in the sources classified as AGB stars, or it could be indicative of some AGB sources
being present in the BGPS catalog.  Indeed, the dust around some AGB and post-AGB stars is
detected at millimeter wavelengths (e.g.~ Buemi et al.~2007; Dehaes et al.~2007;
Ladjal et al.~2010) and it is possible that some BGPS sources are actually
evolved stars.  
The RMS team has performed follow up observations and has identified the
sources in their catalog which are AGB stars.  Although the number of RMS ``Evolved stars''
matched with a BGPS source is consistent with the number expected due to chance 
alignments, we can use the matching statistics to estimate the number of 
GLIMPSE R08 AGB stars expected to be matched with a BGPS source.  
If we assume the matching statistics 
for the RMS AGB stars within the ``Evolved stars'' category 
(12 of 133 were matched with a BGPS source, 9.0\%) also 
apply to the R08 AGB stars, then we expect 296 R08 AGB stars to be matched with BGPS 
sources.  We find a total of 494 R08 AGB stars are matched with BGPS sources, 198 
more than expected based on the matching statistics of the RMS AGB stars. 
If the RMS statistics are representative of the R08 AGB stars as well, then 
roughly 60\%\ of the matched R08 AGB stars are likely to be AGB stars and 40\%\ are 
likely to be YSOs.  
Further observations are required to determine
the nature of the AGB sources which are matched with BGPS sources.
A future paper 
will further 
explore the matching statistics in order to assess the validity
of the magnitude and color cuts made to define the YSO GLIMPSE
Red Source list presented in R08.

Finally, the additional GLIMPSE source catalog contains 21\% of sources that are matched with mm emission (almost three times the level expected from chance alignments), indicating that it does contain a fraction of genuinely young stars.

\subsection{Star Formation Activity in BGPS Sources}\label{bgpssfresults}
In columns $6-9$ of Table \ref{matchstats}, 
we present the detailed numbers/fractions of BGPS sources containing each type 
of mid-IR source.  We assume the following mid-IR source types indicate 
star formation within the matched BGPS source:  GLIMPSE R08 YSOs, sAGB, and xAGB 
stars; GLIMPSE additional sources; EGOs; young RMS source types including 
``Young/old star'', ``HII Region'', ``HII/YSO'', and ``YSO''.  As previously
mentioned (section \ref{matchstatssection}), we use the R08 AGB catalogs
as an indicator of star formation because of the approximate separation of
the YSOs and AGB sources and the resulting contamination of the R08 AGB catalogs 
by YSOs  (see section \ref{redsources}).  
Further observations would be required 
to determine if R08 AGB sources matched with BGPS sources are YSOs or AGB stars,
and we note that some AGB stars matched with BGPS sources will actually be AGB stars.

A total of 3,712 BGPS sources were identified as containing signs
of star formation via the alignment of a mid-IR source, with the caveat that some of 
the matched R08 AGB stars are actually AGB stars and not indicative of star formation.  This 
equates to 44\%\ of the entire BGPS catalog, although this is a lower-limit
to the true fraction since the mid-IR catalogs do not 
overlap completely 
with the BGPS catalog.  Only the RMS catalog overlaps the BGPS
at $l > 65\degree$, and unlike the GLIMPSE
survey it will only identify the brightest mid-IR sources.  If
we consider only the region where all surveys overlap 
($10\degree < l < 65\degree$), we find a total of
2,457 BPGS sources containing a mid-IR source out of 5,067 BGPS sources
in the inner Galaxy (48\%).  

In order to calculate the true number and fraction of BGPS sources that show
signs of active star formation, we must account for the 
occurance of chance alignments.  As previously discussed,
the number of BGPS sources likely to have chance alignments
were calculated for each mid-IR source catalog (the product of columns
6 and 9 in Table \ref{matchstats}).  
We simply subtract the number of BGPS sources identified
as chance alignments from the number of BGPS sources which were
matched with each individual mid-IR source type, and find that 
1,472 of the 8,358 (17.6\%) BGPS sources in the entire catalog are reliable matches. 
If we restrict this calculation to the 5,067 BGPS sources in the area
where all surveys overlap (10\degree$\leq \ell \leq$65\degree), we estimate
that 1,035 (20.4\%) BGPS sources are reliably matched with a mid-IR source. 
These percentages are conservative estimates of the percentage of BGPS sources that
are actively forming stars and are driven so low due to the large
number of BGPS sources expected to be identified via chance alignments
with the additional GLIMPSE sources (1,573 for the entire catalog; 976 for the
overlap area only).

Since we
are unable to identify which mid-IR sources are chance alignments
with our current data set, we consider all matches in the subsequent 
discussion.  Table \ref{bgpscatalog} presents the number of mid-IR
sources within each BGPS source.  Column 1 gives the BGPS source name, 
Columns 2 and 3 give the R.A.~and DEC of the geometric centroid, 
while Columns 4 and 5 give the Galactic coordinates.  The angular radius of 
each BGPS source is given in Column 6.  The number of matched sources from
the EGO, RMS, GLIMPSE R08, and additional GLIMPSE catalogs are given
in Columns 7-10, respectively.  The full table is included in the online 
journal only.

We define a \textit{Confidence} value for each type which is 
the percent of matched BGPS sources which are not due to chance alignments.  
This is defined as the number of BGPS sources with a matching mid-IR source minus 
the number expected from chance alignments, divided by the number of BGPS 
sources with a matching mid-IR source, and is given in column 10 of 
Table \ref{matchstats}.

%Figure 4
\begin{figure*}
\epsscale{0.8}
\plotone{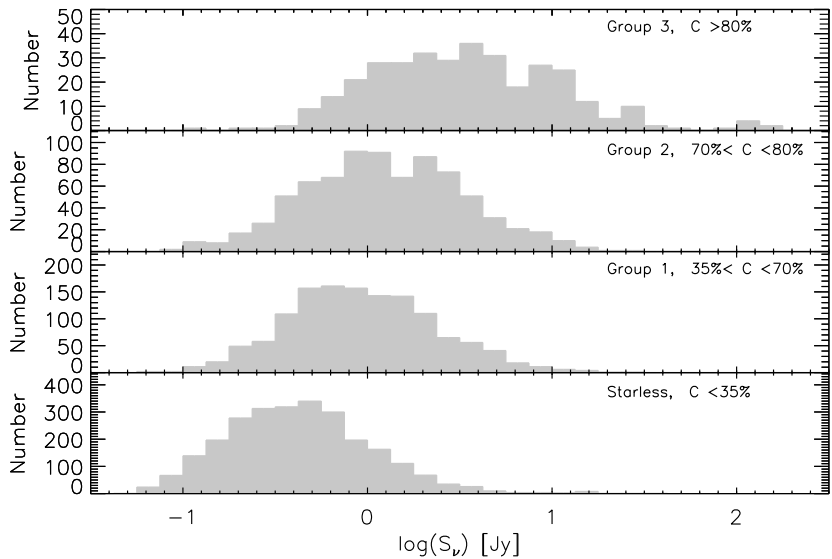} 
\figcaption{\label{fluxvsconf} 
Number of sources as a function of BGPS source flux for the different probability groups of BGPS sources.  Only the 5067 BGPS sources in the region where all catalogs overlap are included.  Note the vertical scale is different for each probability group.  The range of confidence values for each probability group are given in the upper-right corner of each panel.
}
\end{figure*}

This confidence value allows us to estimate the probability that a match with a particular 
mid-IR source indicates ongoing star formation in the BGPS source. The calculated 
confidence values range from 26\%, for BGPS sources matched with RMS evolved stars, 
to roughly 90\%, for the BGPS sources matched to EGO and young RMS sources. The confidence 
is around 70\% for the GLIMPSE R08 YSOs, around 50\% for the additional GLIMPSE 
sources, and around 50\% and 40\% for the GLIMPSE R08 xAGB and sAGB stars 
respectively.

\section{Discussion}\label{discussion}
In this section, we consider only the 5,067 BGPS sources located in the region
where all catalogs overlap ($10$\degree$\le \ell \le 65$\degree).
In order to compare the properties of the BGPS sources as a function of the star-formation activity they contain, we define a classification scheme based on the confidence
parameter listed in Table \ref{matchstats}. We separate the BGPS sources into four groups 
representing the probability of star formation activity:
\begin{enumerate}
\item[3.] BGPS sources containing at least one EGO or young RMS (Young/old star, HII Region, 
HII/YSO, or YSO) source (corresponding to a confidence of >80\%)
\item[2.] BGPS sources containing no EGO or young RMS source, but at least one GLIMPSE R08 YSO (corresponding to a confidence of ~70\%)
\item[1.] BGPS sources containing no EGO, young RMS source, or GLIMPSE R08 YSO, but at least one GLIMPSE R08 AGB star or additional GLIMPSE source (corresponding to a confidence of 40-50\%). The inclusion of the R08 sources classified as AGB stars may seem counter-intuitive, but they are contaminated by genuine YSOs and are therefore to some extent a star formation indicator.
\item[0.] The remaining BGPS sources, which show no sign of star formation in the above catalogs.  We refer to these sources as ``starless''.   These may in fact be forming stars that are too faint to be detected, or too young and too deeply embedded to be visible in the mid-IR.  
\end{enumerate}
In this classification scheme, a BGPS source is assigned to a group based 
on the highest confidence mid-IR source type it was matched with.  For 
example, a BGPS source matched with an EGO, R08 YSO, and 5 additional
GLIMPSE sources would be assigned to group 3 based on the presence
of the EGO source, while a BGPS source without the EGO, but with 
a R08 YSO and 5 additional GLIMPSE sources would be assigned to group 2
based on the presence of the R08 YSO.  

Probability Groups 0 through 3 are comprised of 2,610, 1,324, 793, and
340 BGPS sources, respectively.
The trend of fewer BGPS sources with increasing
confidence parameter reflects the smaller number of 
mid-IR sources in the catalogs required for each higher 
confidence category.  We note that the actual probability of active star 
formation for each BGPS source is unknown.  The probability we refer to throughout
this paper is a measure of the probability that the mid-IR source is a real
association rather than a chance alignment.

\subsection{Observed Properties}\label{obsprops}
Figure \ref{fluxvsconf} shows the number of BGPS sources as a function of 
the flux density integrated over the entire source for the 
BGPS sources in each probability group.  BGPS
sources in  Group 3 with the highest probability of star formation activity
have the highest mean flux, 8.0 Jy, with a 
standard deviation about the mean of 18.0 Jy.  The mean flux decreases
with  star formation activity probability, with a mean of 0.71 Jy and standard deviation 
of 1.4 Jy for the ``starless'' sources.  The flux distributions are
asymmetric with tails extending to large fluxes.  For example, the distribution of
``starless'' sources has a high flux tail extending to approximately
30 Jy.  

%\subsubsection{Free-Free Contribution}
The 1.1 mm BGPS flux densities could be contaminated by free-free emission
from ionized gas if the BGPS source is coincident with an HII region
or Planetary Nebula, for example.  The trend of increasing flux 
density with probability of star formation could be due to an increase
in the contamination from free-free with increasing probability
of star formation.  To quantify the free-free contamination we have 
compared the BGPS flux densities with integrated intensities from
6 cm radio continuum observations toward RMS sources (Urquhart et al.~2009c).
We have identified 223 radio continuum sources from Urquhart et al.~which 
coincide with 159 unique BGPS sources.  
In cases where multiple radio continuum sources coincide with a BGPS source,
we have simply taken the sum of the individual 6 cm fluxes for comparison
to the 1.1 mm flux.  Assuming the 6 cm
emission is on the optically thin side of the free-free spectrum,
we scale the 6 cm emission to 1.1 mm via 
$\Snu(\rm{ff,1.1 mm}) =\Snu(\rm{6 cm}) (\rm{1.1 mm/60 mm})^{0.1}$ since 
$\Snu(\rm{free-free}) \propto \nu^{-0.1}$.  If the 6 cm
emission is on the optically thick side of the free-free spectrum,
then \Snu(ff,1.1 mm) underestimates the free-free flux density,
as the spectrum increases as $\nu^2$ on the optically thick side
(Mezger \& Henderson 1967; Tielens 2005).

%Figure 
\begin{figure}
\epsscale{1.0}
\plotone{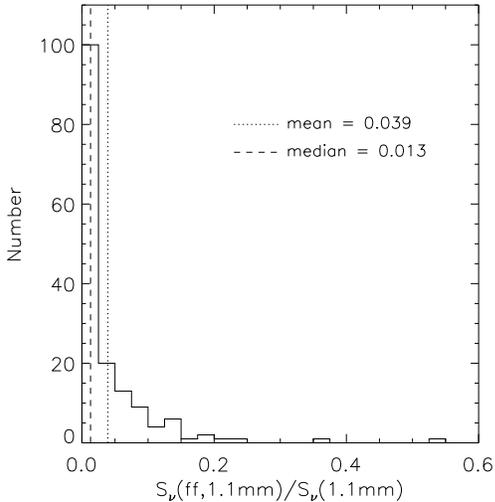}
\figcaption{\label{ffcont}
Number of sources as a function of the fraction of 1.1 mm flux density which is due to free-free emission based on 6 cm radio continuum observations toward RMS sources (Urquhart et al.~2009c) as discussed in section \ref{obsprops}.  The mean fraction of flux density attributed to free-free emission is 0.039, and the median is 0.013.  Overall, free-free emission is not the cause of the trend of increasing flux density seen with increasing probability of star formation, although for some BGPS sources the free-free contribution can be significant ($>50\%$).
}
\end{figure}

Figure \ref{ffcont} shows the distribution of the ratio of the
free-free flux density at 1.1 mm to the 1.1 mm BGPS flux density.  
The dotted line marks the mean fractional contribution from free-free of
$0.039\pm0.066$, and the dashed line marks the median of 0.013.
The minimum and maximum free-free contributions are 0.0002 and 0.54,
respectively, and both values are for BGPS sources in the ``starless''
group.  The RMS 6 cm observations were targeted toward potential
HII regions with the goal of further classifying the RMS sources;
some RMS sources were classified as Planetary Nebulae which we
did not consider as a positive sign of star formation and could 
therefore fall into our ``starless'' probability group.
The vast majority of the BGPS sources including 6 cm emission
fall in probability group 3, with a mean free-free contribution
of 0.033$\pm$0.046, median of 0.012 and a maximum of 0.23.  
For all but four of the 159 BGPS sources with 6 cm radio observations, 
contamination from free-free emission does not exceed 20\%, while 
for the remaining four BGPS sources the free-free is contributing 
up to approximately half of the 1.1 mm flux density.  

Without radio observations for all BGPS 
sources, we are unable to correct for the free-free contamination.  
To assess if the mean level of free-free contamination we see
towards the 159 BGPS sources could be the cause of the increasing
flux density with probability of star formation, we calculate what
fraction of the group 1-3 flux would need to be attributed to 
free-free emission for the mean flux density to match the mean 
of the ``starless'' group.  We find 52\%, 64\%, and 91\%\ for
groups $1-3$, respectively.  This is an order of magnitude higher
than the mean free-free flux density contribution we find with
the RMS 6 cm observations.  Therefore, it is highly unlikely that 
the trend of increasing flux density is due to free-free 
contamination.  However, free-free emission could 
contribute significantly to the flux densities of individual sources. 

The observed increase in flux density with group could also be
a reflection of an increasing dust temperature.  While an increase
in \td\ does explain some of the observed trend in flux density, it
cannot explain the order of magnitude increase in mean flux density
between groups 0 and 3.  We discuss the effects of \td\ in Section
\ref{physicalprops} below.

A SIMBAD search of the 22 ``starless'' 
sources in the inner Galaxy with \Snu $>10$ Jy
revealed that only 4 of the brightest BGPS sources are possibly truly 
starless, while the 
remaining 18 brightest BGPS sources in the inner Galaxy had a combination of  
radio emission and masers as indicators of active star formation. 
This accentuates  that 
while these BGPS sources are ``starless'' in the sense that they do not
contain any mid-IR sources from GLIMPSE or RMS catalogs, they are not
necessarily truly starless sources.  Further comparison to other
surveys will be needed in order to find a truly starless sample
of BGPS sources. 

%Figure n+1
\begin{figure*}
\epsscale{0.8}
\plotone{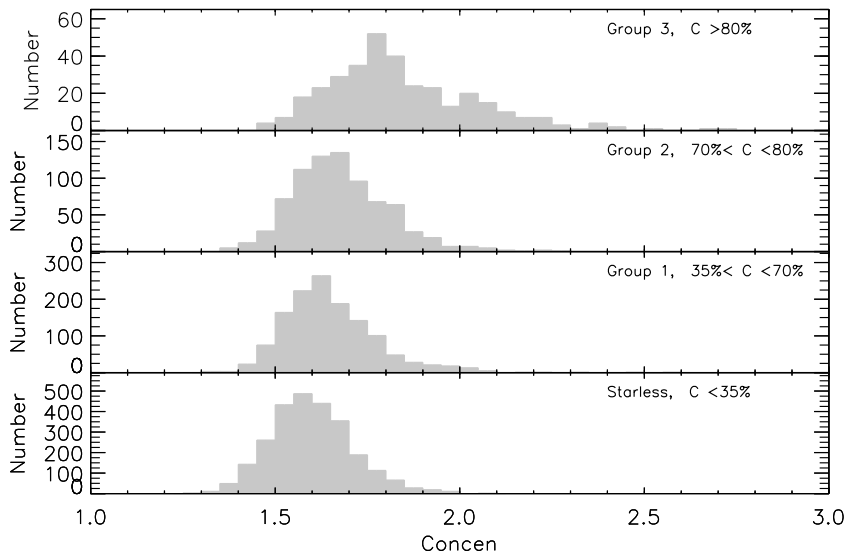} 
\figcaption{\label{concenvsconf} 
Number of sources as a function of BGPS concentration parameter for the different probability groups of BGPS sources.  Only the 5067 BGPS sources in the region where all catalogs overlap are included.  The concentration parameter is given by the ratio of the elliptical radius containing 90\%\ of the millimeter emission to the elliptical radius including 50\%\ of the emission.  A smaller concentration parameter means the source is more compact, while sources with large skirts of emission will have a higher concentration parameter.  Note the vertical scale is different for each probability group.  The range of confidence values for each probability group are given in the upper-right corner of each panel.
}
\end{figure*}

Figure \ref{concenvsconf} shows the number of BGPS sources as a function 
of the concentration parameter in each star formation probability group.  The concentration
parameter is defined as the ratio of the elliptical radius containing
90\% of the total source flux to the elliptical radius containing 
50\% of the total source flux, where the elliptical radius
is given by $R=(R_{maj}R_{min})^{1/2}$.  By this definition, compact sources
will have a low concentration parameter while sources with large
skirts of emission will have a high concentration parameter.
The ``starless'' sources have a lower mean concentration parameter,
and the mean value increases with probability of star formation activity.  The 
majority of the most extended BGPS sources fall within probability Group 3. 
These statistics are likely affected by chance alignments since more extended 
sources are more likely to be identified by chance alignments with the mid-IR
sources.  Since we cannot identify which BGPS-mid-IR source associations are
due to chance alignments, we cannot assess the extent of the effect of chance alignments 
on this trend.  However, we do note that the group 3 sources are unlikely to include
many chance alignments since we expect a very low fraction of matched EGO and RMS sources
to be the result of chance alignments. 

Another possible bias which could affect the statistics of each probability group is
distance.  Due to the limited sensitivity of the mid-IR catalogs, it is possible that 
more distant BGPS sources harbor low-mass YSOs which are not present in the mid-IR
catalogs.  The starless group could include a large number of distant BGPS sources
which are actually forming stars.  

\subsection{Physical Properties}\label{physicalprops}
We create a distance subsample of BGPS sources within the region where all
surveys overlap that have distance estimates from
 a molecular line study of BGPS sources (Schlingman et al.~2011) 
and VLBI parallax measurements (7 BGPS sources; Bartkiewicz et 
al.~2008; Brunthaler et al.~2009; Xu et al.~2009).  
Schlingman et al.~used the VLBI parallax 
based model of Galactic rotation of Reid et al.~(2009) and were able to assign
the near kinematic distance to sources based on the association with a 
GLIMPSE IRDC (Peretto et al.~2009).   Schlingman et al.~additionally
adopted 
distances from the well-studied sample of massive star-forming regions
of Shirley et al.~(2003).  The ``distance subsample''
presented in this section consists of 280 BGPS sources which are located
in the region where all the catalogs overlap.  Since the dominant number of distance
determinations are based on the association of a BGPS source with an
IRDC, the distance subsample is significantly biased 
to the near kinematic distance in the inner Galaxy.  Full results of the
molecular line study are formally presented in 
Schlingman et al.~(2011).

%Figure
\begin{figure*}
\epsscale{0.8}
\plotone{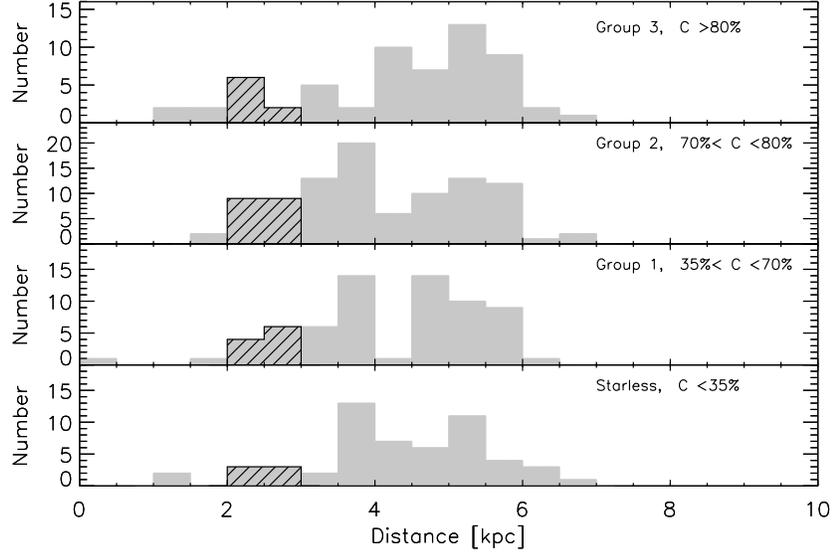}
\figcaption{\label{distance}
Number of sources as a function of distance for each of the different probability groups of BGPS sources which have distance measurements and are located in the region where all catalogs overlap.  The striped histogram represents the 2 kpc subsample.  Note the vertical scale is different for each probability group.  The range of confidence values for each probability group are given in the upper-right corner of each panel.
}
\end{figure*}

%Figure 
\begin{figure*}
\epsscale{0.8}
\plotone{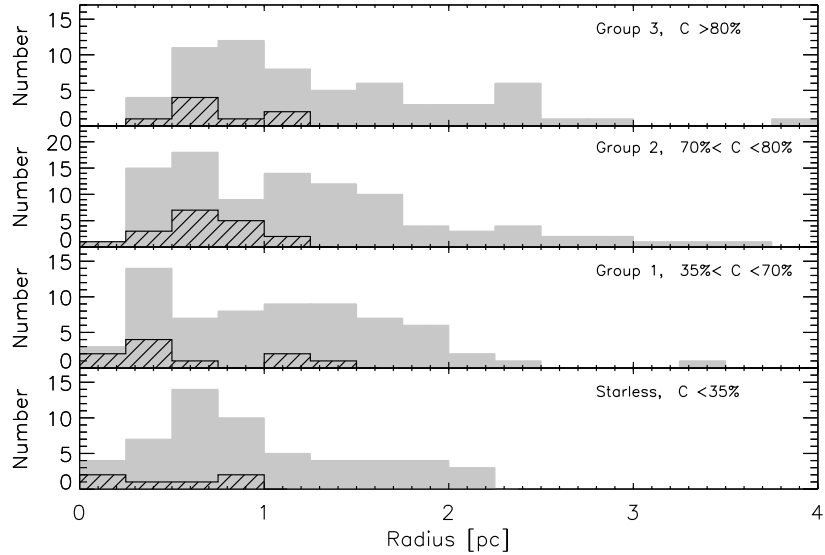}
\figcaption{\label{radius}
Number of sources as a function of physical radius for each of the probability groups of BGPS sources which have distance measurements and are located in the region where all catalogs overlap.  The striped histogram represents the 2 kpc subsample.    The range of confidence values for each probability group are given in the upper-right corner of each panel.
}
\end{figure*}

We calculate properties such as mass, volume-averaged density, column 
density, etc., for the 280 BGPS sources with kinematic distances.  
The distance subsample samples the different probability groups 
well with 55, 67, 97, and 61 from groups 0-3, 
respectively.  Figure \ref{distance} plots a histogram of the distances
as a function of probability of star formation activity.  
The distances
range from 0.21 to 7.0 kpc with no obvious trends with star formation 
probability (see Table \ref{stats}).
Since the BGPS will be sensitive to different structures over such a large 
range in distances, we define a subsample of sources with  $2$ kpc$<d<3$ kpc
(hereafter refered to as the 2 kpc subsample) in order to separate trends 
from the ambiguity introduced by a large 
range of distances.  We have chosen this range of distances because 
Dunham et al.~(2010) find that at distances greater than approximately
2 kpc the BGPS is sensitive to structures most closely related to clumps.
Thus, we expect the 2 kpc subsample to have properties in the typical 
range of clumps (e.g. Table 1 of Bergin \& Tafalla 2007).
We find 6, 10, 18, and 8 sources with distances
of $2-3$ kpc for probability groups 0-3, respectively.  

Figure \ref{radius} plots the number of objects versus physical 
radius for each probability group.  The gray histograms show the 
distance subsample, while the striped histograms mark the 2 kpc 
subsample.  The elliptical radius was determined
by the BGPS source extraction software based on the emission weighted 
moments of the BGPS maps (Rosolowsky et al.~2010).  Radii of 
unresolved sources have been set to an upper limit of half the beam size,
16.5\as.  The physical radii of source in the distance subsample
range from 0.07 to 3.8 pc spanning the
range of sizes expected for the entire hierarchical structure in GMCs
including clouds, clumps, and cores (radius $= 1-7.5$ pc, $0.15-1.5$ 
pc and $0.01-0.1$ pc, respectively; Bergin \& Tafalla 2007).  
The radii of the 2 kpc subsample range from $0.18-1.4$ pc,
and span the range of typical clump radii (see Table \ref{stats}). 
The physical radius increases with star formation probability group
from 0.95 pc for the ``starless'' sources to 1.3 pc for group 3
(see Table \ref{stats}).

%Figure
\begin{figure*}
\epsscale{0.8}
\plotone{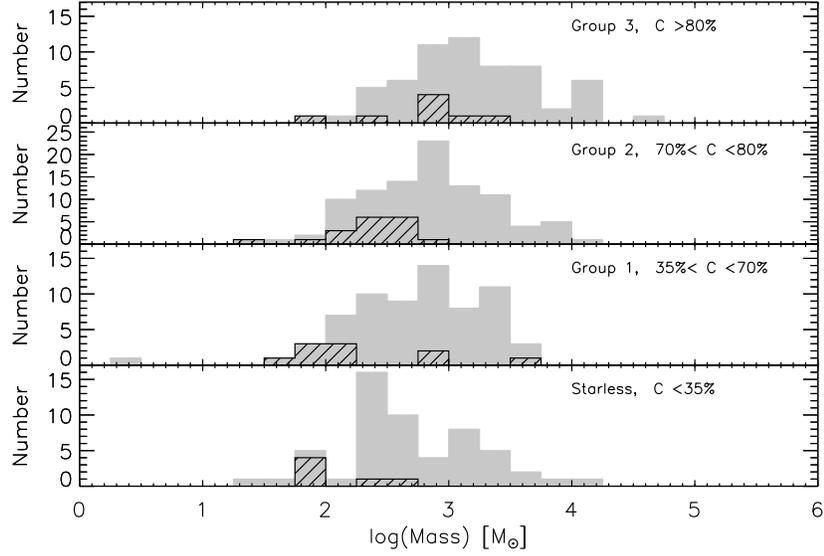}
\figcaption{\label{mass}
Number of sources as a function of isothermal mass for each of the probability groups of BGPS sources which have distance measurements and are located in the region where all catalogs overlap.  The striped histogram represents the 2 kpc subsample.  Note the vertical scale is different for each probability group.    The range of confidence values for each probability group are given in the upper-right corner of each panel.
}
\end{figure*}

We calculate the isothermal dust mass, \miso, assuming a single 
dust temperature, \td, as 
\begin{align} M_{iso}&=\frac{S_{\nu}D^{2}}{\kappa_{\nu}B_{\nu}(\td)}\notag\\
=13.1\ \mbox{M$_\odot$}\ \left(\frac{S_{\nu}}{1\ Jy}\right)&\left(\frac{D}{1\ kpc}\right)^2\left(\frac{e^{13.0\ K / \td}-1}{e^{13.0\ K / 20.0\ K}-1}\right) ,
\label{miso}
\end{align}
where $S_{\nu}$ is the flux density, $D$ is the distance, $\kappa_{\nu}$
is the dust opacity per gram of gas and dust including a gas-to-dust
ratio of 100 ($\kappa_{\nu} = 0.0114$
cm$^2$ g$^{-1}$ at 271.1 GHz; logarithmically interpolated from
OH5 dust from column 5 of Table 1 in Ossenkopf \& Henning 1994), 
and $B_{\nu}$ is the Planck function evaluated at \td. 
The right hand side of the equation is evaluated for \td$=20$ K
based on \ammonia\ observations of BGPS sources in the Gemini OB1 molecular cloud
where a mean gas kinetic temperature of 20 K was found (Dunham et al.~2010).

Although 20 K is a reasonable value of \td\ to assume if no other information is
available, we have both gas kinetic temperature and dust temperature measurements
from other observations.  
M.~K.~Dunham et al.~(in preparation) have surveyed 716 BGPS sources in the
inner Galaxy in \ammonia(1,1), (2,2), and (3,3) and have provided a measurement of
\tk\ towards the 408 BGPS sources detected in \ammonia.  They find \mean{\tk} of
14.0$\pm$3.2 K, 14.8$\pm$3.9 K, 16.1$\pm$4.3 K, and 22.0$\pm$5.9 K based on
130, 143, 102, and 33 BGPS sources from groups $0-3$, respectively.  
For this work we assume the mean gas kinetic temperatures from the \ammonia\ study for
each group. 
By assuming the gas kinetic temperture is equivalent to the dust temperature,
we are assuming that the density is high enough to bring the temperatures into
equilibrium via collisions.  This assumption holds in the dense regions where
the dust effectively shields the gas from the interstellar radiation field (ISRF)
and the gas is heated via collisions with the dust (e.g.~Goldreich \& Kwan 1974;
Evans et al.~2001).

The number of BGPS sources of each \miso\ are plotted for each probability
group in Figure \ref{mass}, where the gray histogram represents
the entire distance subsample and the striped histogram denotes
the 2 kpc subsample.  There is a clear trend of 
increasing mass with increasing probability of star formation activity; 
the median masses in groups $0-3$ are 340, 690, 680, and 1160 \msun, 
respectively.  Similarly, we find a median mass of 52, 94, 170, 710 \msun\ 
for groups $0-3$ of the 2 kpc subsample, respectively.  
The median mass is a more accurate representation of
each group than the mean mass due to the few extremely
high mass ($10^4$ \msun) BGPS sources in groups 0, 2, and 3
(see Table \ref{stats}).  

There are two effects that could artificially produce the observed
trend in mass:  contamination by free-free emission and the assumed
\td.  As discussed previously in section \ref{obsprops}, the free-free emission is not a
significant contribution to the flux densities of most BGPS sources
and is unlikely to cause the trend of increasing mass.  

Here we explore what effects different dust temperatures would have on the observed
trend in mass.  We calculate the temperature required for the median
mass of each group to equal the median mass of group 1 
(where we assume \td$=14.8$ K).  We have chosen
to use the median mass of group 1 as the standard since it is very similar
to the median for group 2 (690 and 680 \msun, respectively).
Dust temperatures of 9.2 K, 15.9 K, and 33.2 K would bring the median
of groups 0, 2, and 3 into alignment with group 1, respectively.  While
these temperatures are reasonable, the values for groups 0 and 3 lie outside of the standard 
deviation of the mean gas kinetic temperatures derived from the \ammonia\ 
observations discussed above.  The \ammonia\ observations are sensitive to
gas kinetic temperatures well above 33 K, so the lower mean temperature
observed toward group 3 sources is not the result of an observational bias.

Thus, 
the trend to higher masses with probability of star formation 
activity is likely real.
\miso\ ranges from 2.4 \msun\ to $3.6\times10^4$ \msun\ for the 
distance subsample, and 19 \msun\ to $2.5\times10^3$ \msun\ for
the 2 kpc subsample.  Again, the distance subsample spans the
range of masses for cores, clumps, and clouds 
($0.5-5$ \msun, $50-500$ \msun, and $10^3-10^4$ \msun, respectively;
Bergin \& Tafalla 2007) while the 2 kpc
subsample encompasses the mass range for clumps without
extending into the range for clouds or cores.

%Figure
\begin{figure*}
\epsscale{0.8}
\plotone{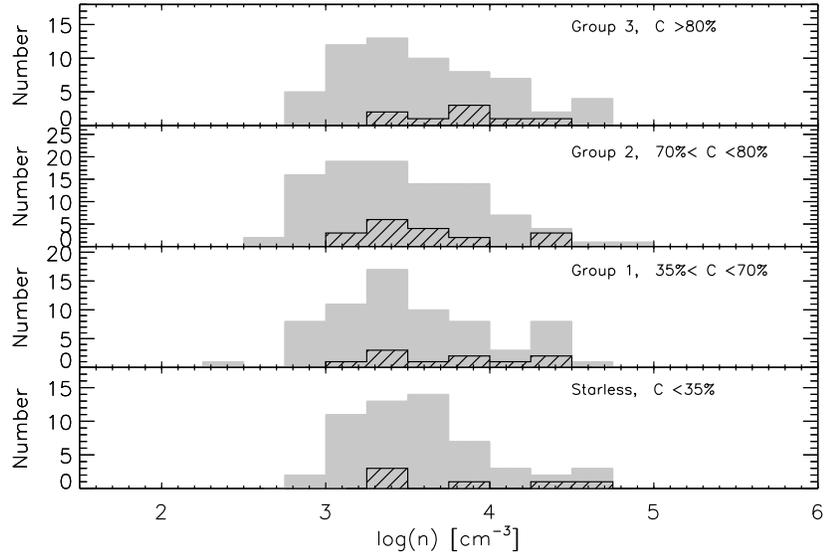}
\figcaption{\label{density}
Number of sources as a function of mean volume-averaged density for the subset of each BGPS source probability group which have distance measurements and are located in the region where all catalogs overlap.  The striped histogram represents the 2 kpc subsample.  Note the vertical scale is different for each probability group.    The range of confidence values for each probability group are given in the upper-right corner of each panel.
}
\end{figure*}

%Figure
\begin{figure*}
\epsscale{0.8}
\plotone{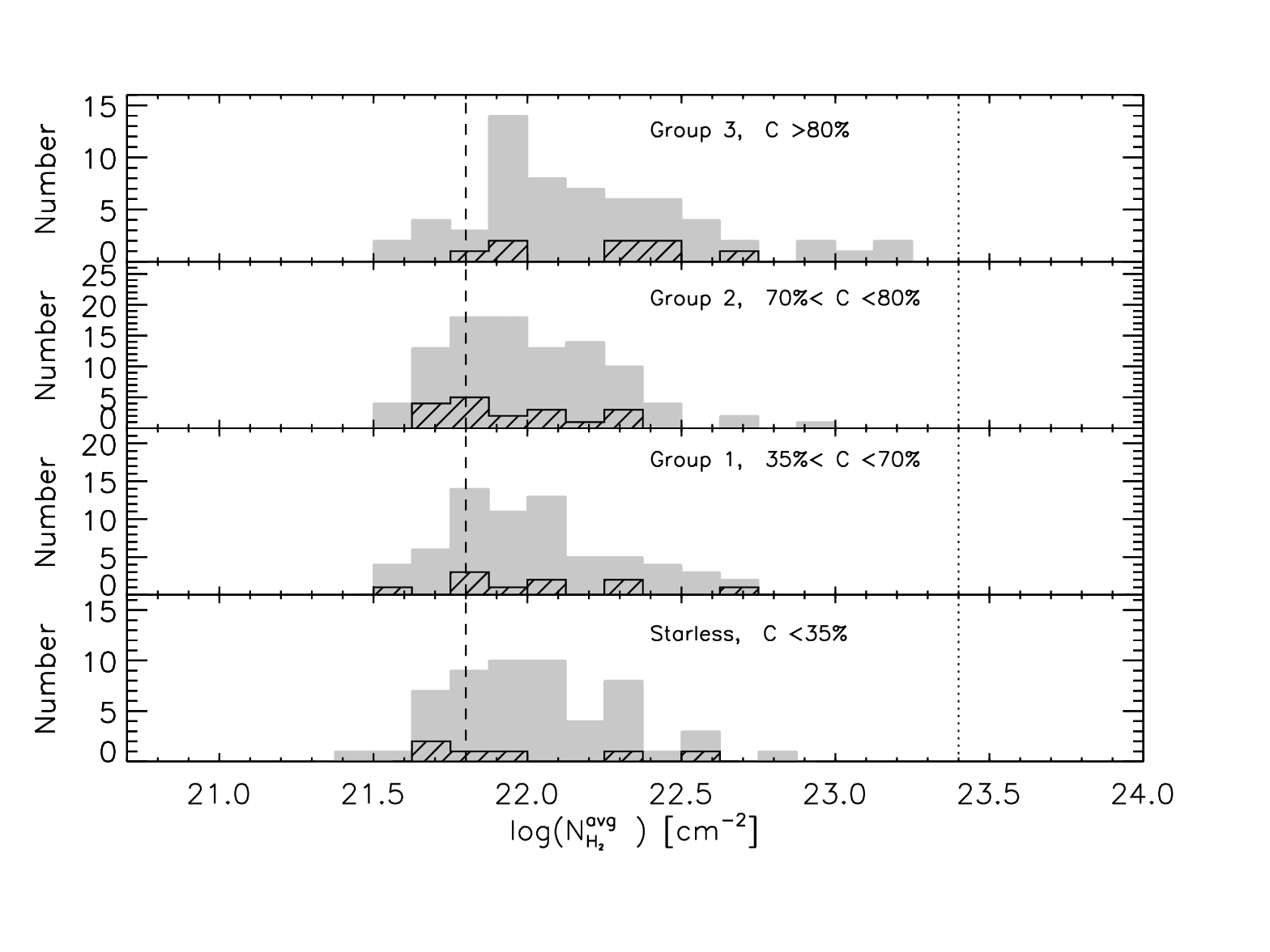}
\figcaption{\label{colden}
Number of sources as a function of H$_2$ column density for the subset of each BGPS source probability group which have distance measurements and are located in the region where all surveys overlap.  The striped histogram represents the 2 kpc subsample.  The dotted line denotes 1 g \cmc, which is the theoretical minimum surface density needed to prevent fragmentation and allow a massive star to form (Krumholz \& McKee 2008).  The dashed line denotes 122.5 \msun\ pc$^{-2}$ which is the threshold column density above which star formation occurs that has been observed in nearby low-mass star-forming regions (Lada et al.~2010; Heiderman et al.~2010).  Note the vertical scale is different for each probability group.  The range of confidence values for each probability group are given in the upper-right corner of each panel.
}
\end{figure*}

From \miso\ we calculate the mean particle density, \n, for each
source in the distance subsample via $\n=3\miso/4\pi R^3$, where 
$R$ is the source radius determined by the BGPS source
extraction algorithm.  This density calculation assumes a constant
density and the effect is a volume-averaged density that is more
reflective of the lower density outer regions of the BGPS source
than the higher density regions.  Figure \ref{density} shows histograms of \n\ 
for each probability group.  All groups span 
almost the same range from $320$ to $5.6\times10^4$ \cmv.  However,
the BGPS sources in groups 0 and 3 are, on average, higher density than those
in groups 1 and 2 (see Table \ref{stats}).  The ``starless'' sources
are less massive but also smaller than BGPS sources in groups $1-3$ and therefore
have densities similar to those of group 3.
The 2 kpc subsample only extends to about $10^3$ \cmv\ on 
the low density end of each distribution, but does extend 
to $10^4$ \cmv\ on the high density end, even in the ``starless'' category.  
Clouds typically range from $50-500$ \cmv, clumps from $10^3-10^4$ \cmv,
and cores from $10^4-10^5$ \cmv\ (Bergin \& Tafalla 2007).  Overall, the distance subsample
spans the entire range of densities, but the 2 kpc subsample
only spans the density ranges for clumps and cores.

%Figure 12
\begin{figure*}
\epsscale{0.8}
\plotone{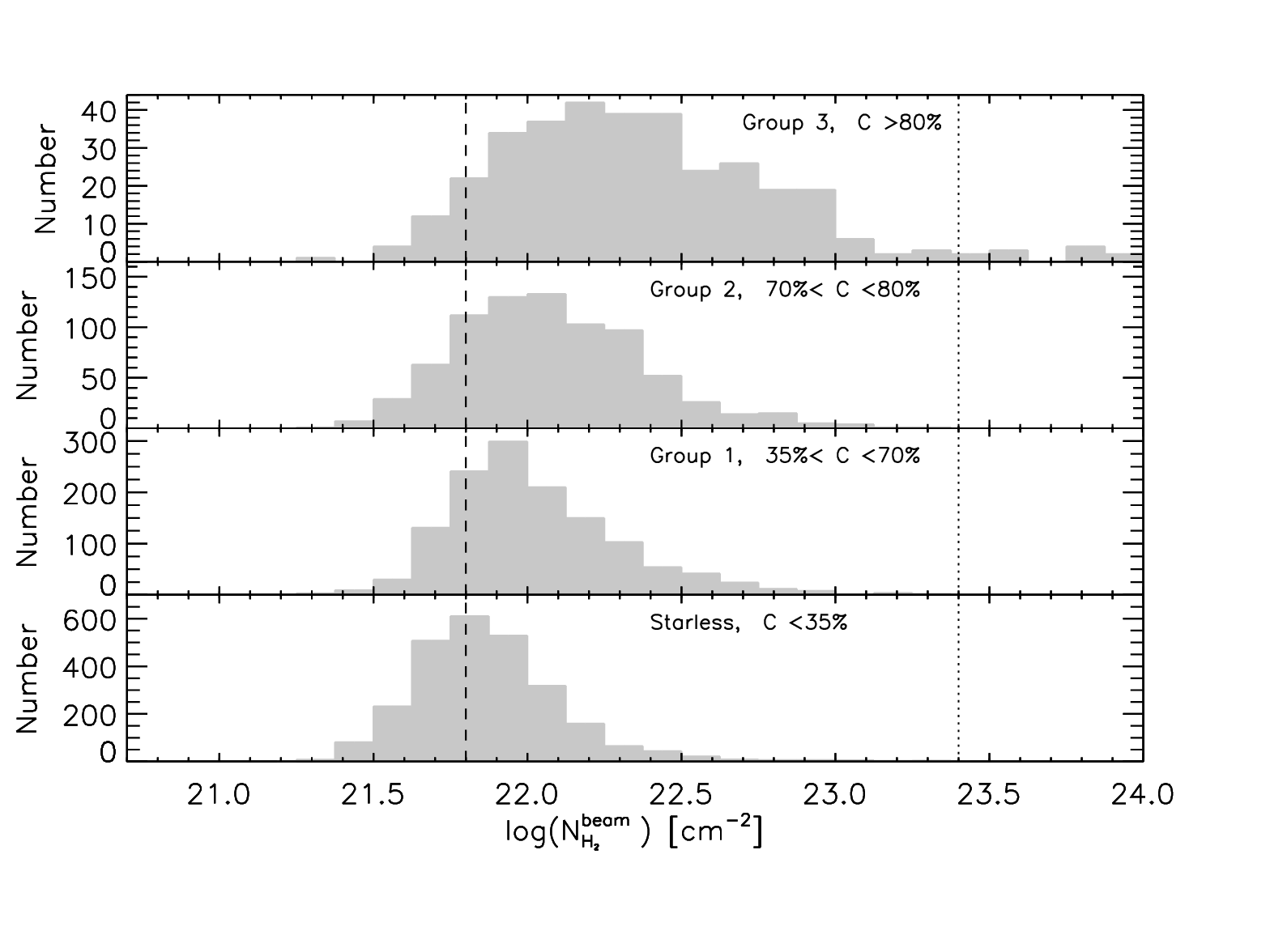}
\figcaption{\label{coldenbeam}
Number of sources as a function of H$_2$ column density per beam for the subset of each BGPS source probability group for all BGPS sources within the region where all catalogs overlap ($10\degree \le \ell \le 65.5$\degree).  The dotted line denotes 1 g \cmc, which is the theoretical minimum surface density needed to prevent fragmentation and allow a massive star to form (Krumholz \& McKee 2008).  The dashed line denotes 122.5 \msun\ pc$^{-2}$ which is the threshold column density above which star formation occurs that has been observed in nearby low-mass star-forming regions (Lada et al.~2010; Heiderman et al.~2010).  Note the vertical scale is different for each probability group.  The range of confidence values for each probability group are given in the upper-right corner of each panel.
}
\end{figure*}

The column density of H$_2$ is given by $N_{H_2}^{avg}=\miso/4\pi \mu m_H R_{obj}^2$,
where $\mu=2.37$ and $m_H$ is the mass of hydrogen.  
Figure \ref{colden} shows the number of sources as a function of
$N_{H_2}^{avg}$ for each probability group in the distance subsample 
(gray histogram) as well
as for the 2 kpc subsample (striped histogram).  There is an 
increase in mean $N_{H_2}^{avg}$ between groups 0$-$2 and
Group 3.  Groups $0-2$ have mean
$N_{H_2}^{avg}$ values near $1.2\times10^{22}$ \cmc\ while
Group 3 has a mean H$_2$ column density of $2.4\times10^{22}$ \cmc (see
Table \ref{stats}).
The dashed line denotes the threshold column density of about 123 \msun\ pc$^{-2}$
which has been observed toward nearby, low-mass star-forming regions
(Lada et al.~2010; Heiderman et al.~2010).  
The dotted line denotes the theoretical
minimum surface density, 1 g \cmc, required to prevent fragmentation and allow 
a massive star to form (Krumholz \& McKee 2008).  
Similar to the volume density, $N_{H_2}^{avg}$ is also
an average quantity.  There are likely higher surface density regions within
the BGPS source where the YSOs are forming, but the BGPS is 
sensitive to the lower density outer regions of the clump (Dunham
et al.~2010).  This is consistent with
the picture that RMS and EGO sources are high-mass YSOs even though
the average surface density is not above the 1 g \cmc\ requirement to prevent
fragmentation.  

Alternatively, we can calculate the H$_2$ column density per beam for all BGPS sources
within the area where all catalogs overlap since this quantity is independent of distance.  The H$_2$ column density per beam is given by 
\begin{equation}
N_{H_2}^{beam}=\frac {\Snu(40\as)}{\Omega_{beam} \mu_{H_2} m_H \kappa_{\nu} B_{\nu}(\td)},
\end{equation}
where \Snu(40\as) is the 1.1 mm flux density within an aperature with a diameter of 40\as, 
$\Omega_{beam}$ is the 
solid angle of the beam, $\mu_{H_2}=2.8$, $\kappa_{\nu}$ is the dust opacity and $B_{\nu}(\td)$
is the Planck function evaluated at the dust temperature, \td, as described above.  We use
\Snu(40\as) as a measure of the flux within a beam since a top-hat function with a 20\as\ radius
has the same solid angle as a Gaussian beam with a FWHM of 33\as.  We have applied an 
aperture correction of 1.46$\pm$0.05 to \Snu(40\as) to account for power that falls outside of 
the 40\as\ aperture due to the sidelobes of the CSO beam (Aguirre et al.~2011).
We calculate this property for all BGPS sources within the region where all surveys overlap
since $N_{H_2}^{beam}$ is independent of distance.  Figure \ref{coldenbeam} shows the distribution of
H$_2$ column density per beam for the BGPS sources in the overlap region for each
probability group.  As above, the dotted line denotes the column density required
to prevent fragmentation (Krumholz \& McKee 2008) and the dashed line
denotes the threshold for effecient star formation (Lada et al.~2010; Heiderman et al.~2010).
The mean column density per beam increases with probability group with
\mean{N_{H_2}^{beam}}$=0.95\times10^{22}$, $1.4\times10^{22}$, $1.6\times10^{22}$,
and $4.7\times10^{22}$ \cmc\ for groups $0-3$, respectively.  The median $N_{H_2}^{beam}$
of group 0 ($7.1\times10^{21}$ \cmc) is similar to the threshold for efficient star
formation ($6.5\times10^{21}$ \cmc; dashed line in Figure \ref{coldenbeam}; Lada et al.~2010;
Heiderman et al.~2010) such that roughly half of the ``starless'' BGPS sources have 
a peak column density high enough to efficiently form stars.  It may be that these sources
appear ``starless'' in this study because of the sensitivity limits of the mid-IR catalogs.
For example, low-mass YSOs within a BGPS source at 5 kpc would not be 
detected by the GLIMPSE surveys.  Further comparisons with other tracers of star formation 
are necessary to determine which sources are truly starless.

Overall, we find that \Snu(int), $R$, \miso, $N_{H_2}^{beam}$, and $N_{H_2}^{avg}$ are all higher 
in BGPS sources actively forming stars.   
Sources identified as actively forming stars
are typically extended with large skirts of emission, 
while the ``starless'' sources are more compact, although this trend is
likely affected by chance alignments since more extended BGPS sources are more
likely to be matched with a mid-IR source by chance.  BGPS sources with 
EGO or RMS sources (group 3) are more massive and have higher volume-averaged
densities than BGPS sources with GLIMPSE sources.
As expected due to the large range of distances, these physical properties 
show that the BGPS sources span the entire hierarchical structure in 
GMCs (clouds, clumps, and cores).  Based on the physical properties 
presented, we confirm that the 2 kpc subsample is mostly comprised of clumps.

\section{Summary}\label{summary}
We have cross-matched the Bolocam Galactic Plane Survey (BGPS)
catalog with available mid-infrared Galactic plane catalogs to determine which BGPS
sources show signs of active star formation.  The mid-infrared
catalogs include the GLIMPSE Red Source catalog (R08; Robitaille et al.~2008),
the \textit{Extended Green Object} catalog (EGOs; Cyganowski et al.~2008),
and the Red \textit{MSX} Source catalog (RMS; Hoare et al.~2004).  
We have also created an additional GLIMPSE Red Source catalog with
a less stringent color selection than was used by R08, namely
$[4.5]-[8.0]>0.75$.  

Overall, 3,712 BGPS sources (44\%) include at least one
mid-IR source.  Within the area where all surveys overlap 
($10\degree < \ell < 65\degree$) 2,457 of 5,067 (48\%) 
BPGS sources include at least one mid-IR source.  
If we account for the number of expected chance 
alignments between the mid-IR and BGPS catalogs, we estimate the 
total number of BGPS sources showing mid-IR signs of active star formation
to be 1,472 over the whole survey (17.6\%) and 1,035 within the area where all
surveys overlap (20.4\%).
These low percentages are conservative lower-limits, while the true
fraction of BGPS sources forming stars is likely higher.  

We additionally divide the BGPS sources into four groups representing
the probability of star formation activity
based on the highest confidence parameter mid-IR source matched
to each BGPS source in order to
study physical  properties as a function of star formation probability.
The sources with the highest probability of star formation activity 
are BGPS sources with an EGO or young RMS source (90\%\ 
confidence).  
The lowest probability group corresponds to sources which were not matched with any
mid-IR sources and are considered to be ``starless'' in this picture.

The mean BGPS flux increases with probability of star formation activity,
with the BGPS sources matched to EGOs and young RMS sources
statistically having the highest fluxes.  The distribution of
``starless'' sources has the lowest mean flux but does include
a tail up to $\sim$30 Jy.  While BGPS sources without mid-IR associations 
are designated ``starless''
here, a literature search shows that these bright sources
do actually show signs of active star formation in other tracers, such 
as masers, X-ray emission and radio emission, which demonstrates
the importance of including other star formation tracers when 
defining a truly starless sample.  For example, these ``starless''
sources may harbor young, highly embedded YSOs which only exhibit
emission at wavelengths longward of 24 \um.  A future paper will
compare the BGPS catalog with the \textit{Spitzer} MIPSGAL catalog of
24, 70, and 160 \um\ sources.  Additionally, this literature search suggests
that the fraction of BGPS sources with active star formation 
is likely higher than our conservative lower-limit of 18\%.

The most compact BGPS sources are typically found to be ``starless'', while
BGPS sources with the highest probability of active star formation are
typically more extended with large skirts of emission.

We define a sample of 280 sources with kinematic distance determinations
ranging from 0.21 to 7.0 kpc, which are strongly biased to
the near distance in the inner Galaxy.  Since BGPS emission
arising from 7 kpc will trace a different average density of gas than that
arising from 1 kpc, we define a subsample of sources with distances
between 2 and 3 kpc.  We see a trend of increasing mass
with increasing probability of star formation activity, 
both in the distance sample and the 2 kpc subsample.  
%The
%highest probability group (with EGO or RMS sources) are on average
%higher density.  
Similarly, there is a trend in H$_2$ column density,
with the highest mean column density found in the group with the highest 
probability of star formation.

The physical properties presented
show that the BGPS sources span the entire hierarchical structure in 
GMCs (clouds, clumps, and cores), as expected due to the large range of
distances.  We also confirm that the 2 kpc subsample is mostly comprised of clumps.
On average, 
BGPS sources matched with EGO or young RMS sources
are more massive and denser than BGPS sources matched with 
GLIMPSE red sources or not matched with any mid-IR sources.

\acknowledgments
The authors thank the anonymous referee for insightful comments which
have improved this work.  We also thank C. Battersby for providing
detailed comments.
T.~Robitaille was supported through the University of Texas at Austin
Tinsley Visiting Scholar program and also by NASA through the Spitzer Space
Telescope Fellowship Program, through a contract issued by the Jet
Propulsion Laboratory, California Institute of Technology under a
contract with NASA.  The BGPS is supported by the National Science 
Foundation through NSF grant AST-0708403.  M.~K.~Dunham and N.~J.~Evans 
were supported by NSF grant AST-0607793 to the University of Texas at 
Austin.  M.~K.~Dunham was additionally supported by a grant from the 
National Radio Astronomy Observatory (NRAO) Student Observing Support
Program, award number GSSP09-0004.  The NRAO is a facility of the 
National Science Foundation, operated under cooperative agreement by 
Associated Universities, Inc.  C.~J.~Cyganowski acknowledges support from NSF 
grant AST-0808119.  C.~J.~Cyganowski was partially supported during this work by 
a NSF Graduate Research Fellowship, and is currently supported by an
NSF Astronomy and Astrophysics Postdoctoral Fellowship under
award AST-1003134.  This paper made use of information from the
Red \textit{MSX} Source (RMS) survey database at www.ast.leeds.ac.uk/RMS which
was constructed with support from the Science and Technology Facilities
Council of the UK.
This research has made use of NASA Astrophysics Data System (ADS) Abstract 
Service and of the SIMBAD database, operated at CDS, Strasbourg, France.

%%%%%%%%%%%%%% References %%%%%%%%%%%%%%%%%%%%%%

\clearpage
%%%%%%%%%%%%%%%%%% Tables %%%%%%%%%%%%%%%%%%%%%
\clearpage
\begin{landscape}
\begin{deluxetable}{lccccccccc}
\tabletypesize{\scriptsize}
\tablewidth{0pt}
\tablecaption{\label{matchstats}Matching Statistics}
\tablehead{
\colhead{}  &\colhead{Objects in} & \multicolumn{3}{c}{Objects Matching BGPS Sources}  &\colhead{BGPS in} & \multicolumn{3}{c}{BGPS Sources Containing Objects} & \colhead{} \\
\cline{3-5}  \cline{7-9}
\colhead{Catalog} & \colhead{Overlap Region} & \colhead{Number} & \colhead{Percent} & \colhead{Randomized \%}  & \colhead{Overlap Region} & \colhead{Number} & \colhead{Percent} & \colhead{Randomized \%} & \colhead{Confidence} 
}
\startdata
GLIMPSE R08 &      &      &        &               & &        &  &      \\
\hfill YSOs & 5053 & 1692 &  33.5  &  8.2$\pm$0.4  & 7360 &  1273 & 17.3 & 5.1$\pm$0.3 & 70.5 \\
\hfill sAGB & 1540 & 169  &  11.0  &  7.2$\pm$0.8  & 7360 &  167  & 2.3 & 1.4$\pm$0.2 &  36.8 \\
\hfill xAGB & 1746 & 325  &  18.6  &  8.8$\pm$0.5  & 7360 &  298  & 4.0 & 1.9$\pm$0.1  & 52.1 \\
GLIMPSE additional & 27278 & 5788 & 21.2 & 7.9$\pm$0.2 & 7360 &  2929 & 39.8 & 21.3$\pm$0.5 & 46.3 \\
EGOs           & 84    & 79   & 90.0 & 8.3$\pm$4.0 & 5067 &  71   & 1.4  & 0.13$\pm$0.05 & 90.8 \\

RMS  \hfill total & 608 & 414       & 68.1 & 8.4$\pm$1.2 & 6062 &  374  & 6.2  &   0.78$\pm$0.11 & 88.6 \\
\hfill Unclassified/Other & 26 & 13 & 50.0 & 7.2$\pm$4.7  & 6062 & 13 & 0.21 & 0.029$\pm$0.019 & 86.5 \\
\hfill Evolved Star      & 133 & 12 & 9.0  &  6.8$\pm$2.2 & 6062 & 12 &  0.20 &  0.14$\pm$0.047 & 26.8 \\
\hfill Young/old star?    & 22 & 16 & 72.7 &  8.8$\pm$6.2 & 6062 & 16 &  0.26 &  0.029$\pm$0.020  & 89.0 \\
\hfill HII Region         & 211 & 180  & 85.3 & 10.4$\pm$2.2 & 6062 & 168 & 2.8 &  0.33$\pm$0.067 & 88.1 \\
\hfill HII/YSO            &  91 & 85 & 93.4 &  8.8$\pm$3.1 & 6062 & 85 & 1.4  & 0.12$\pm$0.041 & 91.6 \\
\hfill YSO                & 124 & 108 & 87.1 &  6.9$\pm$2.3 & 6062 & 106 & 1.7  & 0.13$\pm$0.044 & 92.5 \\
\hline 
                          &     &     &      &              &\hfill Total& 3712& 44.4 &               &   
\enddata 
\end{deluxetable}
\clearpage
\end{landscape}

\begin{deluxetable}{cccccccccc}
\tabletypesize{\scriptsize}
\tablewidth{0pt}
\tablecaption{\label{bgpscatalog}Mid-IR content of the BGPS Sources}
\tablehead{
\colhead{} & \colhead{R.A.\tablenotemark{2}} & \colhead{Dec.\tablenotemark{2}} & \colhead{$\ell$\tablenotemark{2}} & \colhead{b\tablenotemark{2}} & \colhead{Radius\tablenotemark{3}} & \multicolumn{4}{c}{Number Matched} \\ 
\cline{7-10}
\colhead{Source\tablenotemark{1}} & \colhead{(J2000)}  & \colhead{(J2000)}  & \colhead{(\degree)} & \colhead{(\degree)} & \colhead{(\as)} & \colhead{EGO} & \colhead{RMS\tablenotemark{4}} & \colhead{R08} & \colhead{GLIM} 
}
\startdata
G000.000+00.057  &  266.34562  &  -28.89994  &    0.00383  &    0.06324  &   38.4  &   0  &   0  &   0  &   1  \\
G000.004+00.277  &  266.13967  &  -28.78620  &    0.00663  &    0.27649  &   79.6  &   0  &   0  &   0  &   3  \\
G000.006-00.135  &  266.55075  &  -28.99171  &    0.01896  &   -0.13780  &   82.1  &   0  &   0  &   0  &   3  \\
G000.010+00.157  &  266.26410  &  -28.83834  &    0.01913  &    0.15628  &   81.4  &   0  &   0  &   7  &   3  \\
G000.016-00.017  &  266.42551  &  -28.93289  &    0.01216  &   -0.01361  &   93.2  &   0  &   0  &   0  &   6  \\
G000.018-00.431  &  266.83618  &  -29.14625  &    0.01632  &   -0.43125  &  $<$\, 16.5  &   0  &   0  &   0  &   2  \\
G000.020+00.033  &  266.38130  &  -28.90146  &    0.01881  &    0.03579  &   77.2  &   0  &   0  &   0  &   7  \\
G000.020-00.051  &  266.46830  &  -28.94505  &    0.02127  &   -0.05191  &   62.3  &   0  &   0  &   1  &   1  \\
G000.022+00.251  &  266.17683  &  -28.78469  &    0.02494  &    0.24951  &  $<$\, 16.5  &   0  &   0  &   1  &   1  \\
G000.034-00.437  &  266.85490  &  -29.13193  &    0.03704  &   -0.43782  &  $<$\, 16.5  &   0  &   0  &   0  &   0 
\enddata
\tablecomments{The full table is available in the online Journal.}
\tablenotetext{1}{The BGPS source name is based on the Galactic coordinates of the peak of the 1.1 mm emission.}
\tablenotetext{2}{The coordinates of the geometric centroid of the 1.1 mm emission.}
\tablenotetext{3}{For sources unresolved compared to the beam, the radius is set to an upper-limit equal to half the beam size, 16.5\as.}
\tablenotetext{4}{Only young RMS types (``HII Region'', ``HII/YSO'', ``YSO'', and ``Young/old star?'') are included here.}
\end{deluxetable}

\begin{deluxetable}{lcccccc}
\tabletypesize{\scriptsize}
\tablewidth{0pt}
\tablecaption{\label{stats}Physical Property Statistics of BGPS sources in Region Where All Catalogs Overlap}
\tablehead{
\colhead{Property} & \colhead{Probability} & \colhead{} & \colhead{} & \colhead{standard} & \colhead{} & \colhead{} \\
\colhead{(units)} & \colhead{Group} & \colhead{minimum} & \colhead{mean} & \colhead{deviation} & \colhead{median} & \colhead{maximum}
}
\startdata
\Snu      & 0 & 0.06 & 0.71 & 1.4  & 0.41 & 33.1 \\
   (Jy)   & 1 & 0.06 & 1.5  & 1.9  & 0.89 & 27.7 \\
          & 2 & 0.08 & 2.0  & 2.5  & 1.2  & 25.3 \\
          & 3 & 0.13 & 8.0  & 18.0 & 3.4  & 146.8 \\
\cline{1-7}
Concentration   & 0 & 1.3 & 1.6 & 0.11 & 1.6 &  2.1 \\
Parameter       & 1 & 1.3 & 1.7 & 0.13 & 1.6 &  2.6 \\
                & 2 & 1.4 & 1.7 & 0.13 & 1.7 &  2.3 \\
                & 3 & 1.5 & 1.8 & 0.20 & 1.8 &  2.7 \\
\cline{1-7}
$N_{H_2}^{beam}$  & 0 & 0.21 & 0.95 & 1.2 & 0.71 & 23.3 \\
($10^{22}$ \cmc) & 1 & 0.20 & 1.4 & 1.7 & 0.95 & 27.0 \\
                & 2 & 0.22 & 1.6 & 1.6 & 1.1 & 19.3 \\
                & 3 & 0.20 & 4.7 & 10.4 & 2.1 & 99.3 \\
\cline{1-7}
\multicolumn{7}{c}{Distance Subsample}\\
\cline{1-7}
Distance       & 0 & 1.04 & 4.3 & 1.2 & 4.4 & 6.7 \\
   (kpc)       & 1 & 0.21 & 4.2 & 1.2 & 4.5 & 6.4 \\
               & 2 & 1.9  & 4.1 & 1.2 & 3.8 & 7.0 \\
               & 3 & 1.3  & 4.4 & 1.4 & 4.6 & 7.0 \\
\cline{1-7}
Radius      & 0 & 0.12 & 0.95 & 0.55 & 0.86 & 2.0 \\
   (pc)     & 1 & 0.07 & 1.1  & 0.63 & 1.0  & 3.4 \\
            & 2 & 0.18 & 1.2  & 0.75 & 1.1  & 3.5 \\
            & 3 & 0.35 & 1.3  & 0.72 & 1.1  & 3.8 \\
\cline{1-7}
\miso(\td)   & 0 & 19  & 1050 & 1720 & 340 & 1.0$\times10^4$ \\
      (\msun)& 1 & 2.4 & 1040 & 1130 & 690 & 5380 \\
             & 2 & 30  & 1450 & 2110 & 680 & 1.2$\times10^4$ \\
             & 3 & 90  & 3230 & 5450 & 1160 & 3.6$\times10^4$ \\
\cline{1-7}
\n           & 0 & 0.73 & 7.1  & 11.0 & 3.3 & 56.1 \\
($10^3$ \cmv)& 1 & 0.32 & 6.7  & 9.4  & 2.5 & 50.2 \\
             & 2 & 0.41 & 5.8  & 10.7 & 2.5 & 85.7 \\
             & 3 & 0.60 & 7.2  & 9.7  & 3.2 & 45.8 \\
\cline{1-7}
$N_{H_2}^{avg}$    & 0 & 3.1 & 12.8 & 9.7 & 9.6 & 56.3 \\
($10^{21}$ \cmc)  & 1 & 3.7 & 12.8 & 9.9 & 9.7 & 47.8 \\
                 & 2 & 3.2 & 12.4 & 10.8 & 9.2 & 82.2 \\
                 & 3 & 3.4 & 24.2 & 31.3 & 12.6 & 154.0 \\
\cline{1-7}
\multicolumn{7}{c}{2 kpc Subsample}\\
\cline{1-7}
Radius    & 0 & 0.19 & 0.49 & 0.29 & 0.51 & 0.90 \\
 (pc)     & 1 & 0.19 & 0.65 & 0.41 & 0.49 & 1.4 \\
          & 2 & 0.18 & 0.67 & 0.27 & 0.65 & 1.1 \\
          & 3 & 0.41 & 0.74 & 0.23 & 0.73 & 1.1 \\
\cline{1-7}
\miso(\td)   & 0 & 37 & 79  & 56  & 52 & 179 \\
(\msun)      & 1 & 26 & 390 & 767 & 94 & 2520 \\
             & 2 & 19 & 192 & 130 & 170 & 546 \\
             & 3 & 90 & 792 & 609 & 710 & 2120 \\
\cline{1-7}
\n           & 0 & 1.0 & 8.2  & 11.3 & 3.6 & 29.5 \\
($10^3$ \cmv)& 1 & 0.85 & 5.3 & 5.8  & 3.7 & 16.3 \\
             & 2 & 0.91 & 4.0 & 4.2  & 2.3 & 14.2 \\
             & 3 & 2.0 & 9.1  & 6.9  & 8.1 & 22.3 \\
\cline{1-7}
$N_{H_2}^{avg}$         & 0 & 2.7 & 7.2 & 6.8 & 4.7 & 19.9 \\
($10^{21}$ \cmc)  & 1 & 2.4 & 7.9 & 6.9 & 6.1 & 25.4 \\
                 & 2 & 3.1 & 6.4 & 3.5 & 5.0 & 14.5 \\
                 & 3 & 6.3 & 21.0 & 13.7 & 20.4 & 44.6
\enddata
\end{deluxetable}


\begin{thebibliography}{}

\bibitem[Aguirre et al.(2011)]{2011ApJS..192....4A} Aguirre, J.~E., et al.\ 2011, \apjs, 192, 4
\bibitem[Bartkiewicz et al.(2008)]{2008A&A...490..787B} Bartkiewicz, A., Brunthaler, A., Szymczak, M., van Langevelde, H.~J., \& Reid, M.~J.\ 2008, \aap, 490, 787
\bibitem[Benjamin et al.(2003)]{Benjamin:03} Benjamin, R.~A., et al.\ 2003, \pasp, 115, 953 
\bibitem[Bergin \& Tafalla(2007)]{2007ARA&A..45..339B} Bergin, E.~A., \& Tafalla, M.\ 2007, \araa, 45, 339 
\bibitem[Breen et al.(2010)]{2010MNRAS.401.2219B} Breen, S.~L., Ellingsen, S.~P., Caswell, J.~L., \& Lewis, B.~E.\ 2010, \mnras, 401, 2219
\bibitem[Brunthaler et al.(2009)]{2009ApJ...693..424B} Brunthaler, A., Reid, M.~J., Menten, K.~M., Zheng, X.~W., Moscadelli, L., \& Xu, Y.\ 2009, \apj, 693, 424 
\bibitem[Carey et al.(1998)]{1998ApJ...508..721C} Carey, S.~J., Clark, F.~O., Egan, M.~P., Price, S.~D., Shipman, R.~F., \& Kuchar, T.~A.\ 1998, \apj, 508, 721
\bibitem[Carey et al.(2009)]{2009PASP..121...76C} Carey, S.~J., et al.\ 2009, \pasp, 121, 76
\bibitem[Chambers et al.(2009)]{2009ApJS..181..360C} Chambers, E.~T., Jackson, J.~M., Rathborne, J.~M., \& Simon, R.\ 2009, \apjs, 181, 360
\bibitem[Churchwell et al.(2009)]{2009PASP..121..213C} Churchwell, E., et al.\ 2009, \pasp, 121, 213
\bibitem[Churchwell et al.(1990)]{1990A&AS...83..119C} Churchwell, E., Walmsley, C.~M., \& Cesaroni, R.\ 1990, \aaps, 83, 119 
\bibitem[Churchwell(2002)]{2002ARA&A..40...27C} Churchwell, E.\ 2002, \araa, 40, 27 
\bibitem[Cyganowski et al.(2008)]{Cyganowsky:08} Cyganowski, C.~J., et al.\ 2008, \aj, 136, 2391
\bibitem[Cyganowski et al.(2009)]{2009ApJ...702.1615C} Cyganowski, C.~J., Brogan, C.~L., Hunter, T.~R., \& Churchwell, E.\ 2009, \apj, 702, 1615
\bibitem[De Buizer et al.(2005)]{2005ApJS..156..179D} De Buizer, J.~M., Radomski, J.~T., Telesco, C.~M., \& Pi{\~n}a, R.~K.\ 2005, \apjs, 156, 179 
\bibitem[Di Francesco(2008)]{2008AAS...212.9603D} Di Francesco, J.\ 2008, Bulletin of the American Astronomical Society, 40, 271 
\bibitem[Dunham et al.(2010)]{2010ApJ...717.1157D} Dunham, M.~K., et al.\ 2010, \apj, 717, 1157 
\bibitem[Egan et al.(1998)]{1998ApJ...494L.199E} Egan, M.~P., Shipman, R.~F., Price, S.~D., Carey, S.~J., Clark, F.~O., \& Cohen, M.\ 1998, \apjl, 494, L199
\bibitem[Elia et al.(2010)]{2010A&A...518L..97E} Elia, D., et al.\ 2010, \aap, 518, L97 
\bibitem[Ellingsen et al.(2007)]{2007IAUS..242..213E} Ellingsen, S.~P., Voronkov, M.~A., Cragg, D.~M., Sobolev, A.~M., Breen, S.~L., \& Godfrey, P.~D.\ 2007, IAU Symposium, 242, 213 
\bibitem[Evans et al.(2001)]{2001ApJ...557..193E} Evans, N.~J., II, Rawlings, J.~M.~C., Shirley, Y.~L., \& Mundy, L.~G.\ 2001, \apj, 557, 193 
\bibitem[Glenn et al.(2003)]{2003SPIE.4855...30G} Glenn, J., et al.\ 2003, \procspie, 4855, 30
\bibitem[Goldreich \& Kwan(1974)]{1974ApJ...189..441G} Goldreich, P., \& Kwan, J.\ 1974, \apj, 189, 441 
\bibitem[Haig et al.(2004)]{2004SPIE.5498...78H} Haig, D.~J., et al.\ 2004, \procspie, 5498, 78
\bibitem[Heiderman et al. (2010)]{heiderman}Heiderman, A., et al.~2010, ApJ, 723, 1019 
\bibitem[Hoare et al.(2004)]{Hoare:04} Hoare, M.~G., Lumsden, S.~L., Oudmaijer, R.~D., Busfield, A.~L., King, T.~L., \& Moore, T.~L.~J.\ 2004, Milky Way Surveys: The Structure and Evolution of our Galaxy, 317, 156
\bibitem[Krumholz \& McKee(2008)]{2008Natur.451.1082K} Krumholz, M.~R., \& McKee, C.~F.\ 2008, \nat, 451, 1082 
\bibitem[Kurtz et al.(2000)]{2000prpl.conf..299K} Kurtz, S., Cesaroni, R., Churchwell, E., Hofner, P., \& Walmsley, C.~M.\ 2000, Protostars and Planets IV, 299 
\bibitem[Lada et al.(2010)]{2010arXiv1009.2985L} Lada, C.~J., Lombardi, M., \& Alves, J.~F.\ 2010, arXiv:1009.2985
%\bibitem[Lada et al.~(2010)]{Ladasubmitted}Lada, C.~J., Lombardi, M., Alves, J.~F.\ 2010, ApJ, submitted
\bibitem[Longmore et al.(2007)]{2007MNRAS.379..535L} Longmore, S.~N., Burton, M.~G., Barnes, P.~J., Wong, T., Purcell, C.~R., \& Ott, J.\ 2007, \mnras, 379, 535 
\bibitem[Matthews et al.(2009)]{2009AJ....138.1380M} Matthews, H., et al.\ 2009, \aj, 138, 1380 
\bibitem[McKee \& Ostriker(2007)]{2007ARA&A..45..565M} McKee, C.~F., \& Ostriker, E.~C.\ 2007, \araa, 45, 565 
\bibitem[Mezger \& Henderson(1967)]{1967ApJ...147..471M} Mezger, P.~G., \& Henderson, A.~P.\ 1967, \apj, 147, 471
\bibitem[Minier et al.(2005)]{2005A&A...429..945M} Minier, V., Burton, M.~G., Hill, T., Pestalozzi, M.~R., Purcell, C.~R., Garay, G., Walsh, A.~J., \& Longmore, S.\ 2005, \aap, 429, 945 
\bibitem[Molinari et al.(2010)]{2010PASP..122..314M} Molinari, S., et al.\ 2010, \pasp, 122, 314 
%\bibitem[Moscadelli et al.(2009)]{2009ApJ...693..406M} Moscadelli, L., Reid, M.~J., Menten, K.~M., Brunthaler, A., Zheng, X.~W., \& Xu, Y.\ 2009, \apj, 693, 406 
\bibitem[Motte et al.(2003)]{2003ApJ...582..277M} Motte, F., Schilke, P., \& Lis, D.~C.\ 2003, \apj, 582, 277
\bibitem[Motte et al.(2007)]{2007A&A...476.1243M} Motte, F., Bontemps, S., Schilke, P., Schneider, N., Menten, K.~M., \& Brogui{\`e}re, D.\ 2007, \aap, 476, 1243 
\bibitem[Mottram et al.(2007)]{2007A&A...476.1019M} Mottram, J.~C., Hoare, M.~G., Lumsden, S.~L., Oudmaijer, R.~D., Urquhart, J.~S., Sheret, T.~L., Clarke, A.~J., \& Allsopp, J.\ 2007, \aap, 476, 1019 
\bibitem[Mueller et al.(2002)]{2002ApJS..143..469M} Mueller, K.~E., Shirley, Y.~L., Evans, N.~J., II, \& Jacobson, H.~R.\ 2002, \apjs, 143, 469
\bibitem[Peretto \& Fuller(2009)]{2009A&A...505..405P} Peretto, N., \& Fuller, G.~A.\ 2009, \aap, 505, 405 
\bibitem[Peters et al.(2010a)]{2010ApJ...711.1017P} Peters, T., Banerjee, R., Klessen, R.~S., Mac Low, M.-M., Galv{\'a}n-Madrid, R., \& Keto, E.~R.\ 2010a, \apj, 711, 1017 
\bibitem[Peters et al.(2010b)]{2010arXiv1003.4998P} Peters, T., Mac Low, M.-M., Banerjee, R., Klessen, R.~S., \& Dullemond, C.~P.\ 2010b, arXiv:1003.4998 
\bibitem[Price et al.(2001)]{Price:01} Price, S.~D., Egan, M.~P., Carey, S.~J., Mizuno, D.~R., \& Kuchar, T.~A.\ 2001, \aj, 121, 2819
\bibitem[Purcell et al.(2008)]{2008ASPC..387..389P} Purcell, C.~R., Hoare, M.~G., \& Diamond, P.\ 2008, Massive Star Formation: Observations Confront Theory, 387, 389 
\bibitem[Purcell et al.(2009)]{2009MNRAS.394..323P} Purcell, C.~R., Longmore, S.~N., Burton, M.~G., Walsh, A.~J., Minier, V., Cunningham, M.~R., \& Balasubramanyam, R.\ 2009, \mnras, 394, 323
\bibitem[Rathborne et al.(2005)]{2005ApJ...630L.181R} Rathborne, J.~M., Jackson, J.~M., Chambers, E.~T., Simon, R., Shipman, R., \& Frieswijk, W.\ 2005, \apjl, 630, L181  
\bibitem[Rathborne et al.(2006)]{2006ApJ...641..389R} Rathborne, J.~M., Jackson, J.~M., \& Simon, R.\ 2006, \apj, 641, 389 
\bibitem[Rathborne et al.(2007)]{2007ApJ...662.1082R} Rathborne, J.~M., Simon, R., \& Jackson, J.~M.\ 2007, \apj, 662, 1082 
\bibitem[Rathborne et al.(2008)]{2008ApJ...689.1141R} Rathborne, J.~M., Jackson, J.~M., Zhang, Q., \& Simon, R.\ 2008, \apj, 689, 1141 
\bibitem[Rathborne et al.(2010)]{2010ApJ...715..310R} Rathborne, J.~M., Jackson, J.~M., Chambers, E.~T., Stojimirovic, I., Simon, R., Shipman, R., \& Frieswijk, W.\ 2010, \apj, 715, 310 
\bibitem[Reach et al.(2006)]{2006AJ....131.1479R} Reach, W.~T., et al.\ 2006, \aj, 131, 1479 
%\bibitem[Reid et al.(2009a)]{2009ApJ...693..397R} Reid, M.~J., Menten, K.~M., Brunthaler, A., Zheng, X.~W., Moscadelli, L., \& Xu, Y.\ 2009a, \apj, 693, 397
\bibitem[Reid et al.(2009)]{2009ApJ...700..137R} Reid, M.~J., et al.\ 2009, \apj, 700, 137 
%\bibitem[Reid et al.(2009c)]{2009ApJ...705.1548R} Reid, M.~J., Menten, K.~M., Zheng, X.~W., Brunthaler, A., \& Xu, Y.\ 2009c, \apj, 705, 1548
\bibitem[Robitaille et al.(2006)]{2006ApJS..167..256R} Robitaille, T.~P., Whitney, B.~A., Indebetouw, R., Wood, K., \& Denzmore, P.\ 2006, \apjs, 167, 256 
\bibitem[Robitaille et al.(2008)]{Robitaille:08} Robitaille, T.~P., et al.\ 2008, \aj, 136, 2413 
\bibitem[Rosolowsky et al.(2010)]{2010ApJS..188..123R} Rosolowsky, E., et al.\ 2010, \apjs, 188, 123
\bibitem[Schlingman et al.(2011)]{schlingmansubmitted} Schlingman, W., et al.\ 2010, \apj, submitted
\bibitem[Schuller et al.(2009)]{2009A&A...504..415S} Schuller, F., et al.\ 2009, \aap, 504, 415 
\bibitem[Shaver et al.(1983)]{1983MNRAS.204...53S} Shaver, P.~A., McGee, R.~X., Newton, L.~M., Danks, A.~C., \& Pottasch, S.~R.\ 1983, \mnras, 204, 53 
\bibitem[Shirley et al.(2003)]{2003ApJS..149..375S} Shirley, Y.~L., Evans, N.~J., II, Young, K.~E., Knez, C., \& Jaffe, D.~T.\ 2003, \apjs, 149, 375 
\bibitem[Shu et al.(1987)]{1987ARA&A..25...23S} Shu, F.~H., Adams, F.~C., \& Lizano, S.\ 1987, \araa, 25, 23
\bibitem[Simon et al.(2006)]{2006ApJ...653.1325S} Simon, R., Rathborne, J.~M., Shah, R.~Y., Jackson, J.~M., \& Chambers, E.~T.\ 2006, \apj, 653, 1325 
\bibitem[Smith et al.(2006)]{2006ApJ...645.1264S} Smith, H.~A., Hora, J.~L., Marengo, M., \& Pipher, J.~L.\ 2006, \apj, 645, 1264 
\bibitem[Soille et al. 1999]{Soille1999}Soille P., 1999, Morphological Image Analysis, Springer, Berlin
\bibitem[Tielens(2005)]{2005pcim.book.....T} Tielens, A.~G.~G.~M.\ 2005, The Physics and Chemistry of the Interstellar Medium, by A.~G.~G.~M.~Tielens, pp.~.~ISBN 0521826349.~Cambridge, UK: Cambridge University Press,  2005.,  

\bibitem[Urquhart et al.(2007a)]{2007A&A...461...11U} Urquhart, J.~S., Busfield, A.~L., Hoare, M.~G., Lumsden, S.~L., Clarke, A.~J., Moore, T.~J.~T., Mottram, J.~C., \& Oudmaijer, R.~D.\ 2007a, \aap, 461, 11 
\bibitem[Urquhart et al.(2007b)]{2007A&A...474..891U} Urquhart, J.~S., et al.\ 2007b, \aap, 474, 891
\bibitem[Urquhart et al.(2008a)]{2008A&A...487..253U} Urquhart, J.~S., et al.\ 2008a, \aap, 487, 253 
\bibitem[Urquhart et al.(2008b)]{2008ASPC..387..381U} Urquhart, J.~S., Hoare, M.~G., Lumsden, S.~L., Oudmaijer, R.~D., \& Moore, T.~J.~T.\ 2008b, Massive Star Formation: Observations Confront Theory, 387, 381
\bibitem[Urquhart et al.(2009a)]{2009A&A...501..539U} Urquhart, J.~S., et al.\ 2009a, \aap, 501, 539 
\bibitem[Urquhart et al.(2009b)]{2009A&A...507..795U} Urquhart, J.~S., et al.\ 2009b, \aap, 507, 795
\bibitem[Urquhart et al.(2009c)]{2009A&A...501..539U} Urquhart, J.~S., et al.\ 2009c, \aap, 501, 539 
%\bibitem[Wolfire \& Churchwell(1994)]{1994ApJ...427..889W} Wolfire, M.~G., \& Churchwell, E.\ 1994, \apj, 427, 889 
%\bibitem[Xu et al.(2006)]{2006Sci...311...54X} Xu, Y., Reid, M.~J., Zheng, X.~W., \& Menten, K.~M.\ 2006, Science, 311, 54 
\bibitem[Xu et al.(2009)]{2009ApJ...693..413X} Xu, Y., Reid, M.~J., Menten, K.~M., Brunthaler, A., Zheng, X.~W., \& Moscadelli, L.\ 2009, \apj, 693, 413 
\bibitem[Zinnecker \& Yorke(2007)]{2007ARA&A..45..481Z} Zinnecker, H., \& Yorke, H.~W.\ 2007, \araa, 45, 481

\end{thebibliography}
\end{document}